\documentclass[pagesize, titlepage=false, oneside]{scrbook} %,openany
\counterwithin*{chapter}{part} 

\makeatletter
\renewcommand*{\@fnsymbol}[1]{\@arabic{#1}}
\makeatother

\usepackage{style}

\date{} % to avoid printing current dates on articles via \maketitle
\begin{document}

\allowdisplaybreaks
\renewcommand{\thesection}{\arabic{section}}
\setcounter{chapter}{0}

\begin{refsection}
%---------------------------------------------------------------------------------------------------------------------
% -------------------------------------------------
% paper titlepage
% -------------------------------------------------

% Title
%
\title{An Exact Solver for Submodular Knapsack Problems}
% Authors and affiliations
\author{Sabine M{\"u}nch, Stephen Raach}

% Affiliations (by order) separated by \and
\institute{
Trier University, 54286 Trier, Germany \\
\email{\{muench/raach\}@uni-trier.de}
}

{
\maketitle
% \let\newpage\relax
% -------------------------------------------------
\begin{abstract}
We study the problem of maximizing a monotone increasing submodular function over a set of weighted elements subject to a knapsack constraint. 
Although this problem is NP-hard, many applications require exact solutions, as approximate solutions are often insufficient in practice.
To address this need, we propose an exact branch-and-bound algorithm tailored for the submodular knapsack problem and introduce several acceleration techniques to enhance its efficiency.
We evaluate these techniques on artificial instances of three benchmark problems as well as on instances derived from real-world data. We compare the proposed solver with two solvers by~\textcite{sakaue2018}, which currently achieve the strongest performance reported in the literature, as well as with a branch-and-cut algorithm implemented using Gurobi that solves a binary linear reformulation of the submodular knapsack problem, demonstrating that our methods are highly successful.
\end{abstract}

}

\section{Introduction.}\label{e_sec1}
Submodular maximization is a key component of various classical combinatorial optimization problems including graph cuts~\parencite[see, e.g.,][]{goemans1995}, set cover~\parencite[see, e.g.,][]{feige1998}, facility location~\parencite[see, e.g.,][]{cornuejols1977, ageev1999}, and generalized assignment problems~\parencite[see, e.g.,][]{fleischer2011,feige2006,fleischer2006,chekuri2005}.

In addition, submodular maximization plays a crucial role in numerous application domains such as non-parametric learning~\parencite[see, e.g.,][]{mirzasoleiman2016}, sensor placement~\parencite[see, e.g.,][]{krause2007,krause2008}, data subset selection~\parencite[see, e.g.,][]{wei2015}, influence maximization in social networks~\parencite[see, e.g.,][]{kempe2015}, and extractive document and image summarization~\parencite[see, e.g.,][]{lin2011,tschiatschek2014}.

\begin{definition}
	In the following, let $I$ be a finite set of cardinality $n\in \mathbb{N}$, and ${f\colon2^I\rightarrow \mathbb{R}_{\geq 0}}$ be a \textbf{submodular} ($f(X\cup\{i\})-f(X) \geq f(Y\cup \{i\})-f(Y)$ for $X\subseteq Y\subseteq I$, $i \in I\setminus Y$), \textbf{monotone increasing} ($f(X)\leq f(Y)$ for any $X\subseteq Y\subseteq I$), and \textbf{normalized} ($f(\emptyset)=0$) function. 
    We assume that every $i\in I$ is associated with a \textbf{weight} $w_i\in \R_{>0}$, and denote by $w(X)\coloneqq\sum_{i\in X} w_i$ the weight of a set $X\subseteq I$.   
\end{definition}

We consider the classical \textsc{Submodular Knapsack-Constrained Maximization} problem, which is given by 
\begin{align}
    \max \{f(X)\colon w(X) \leq B,\, X\subseteq I\},\label{e_prob1}
\end{align}
where $B\in \R_{\geq 0}$ is the \textbf{knapsack capacity}. 

\begin{definition}
We encode an \textbf{instance} of \emph{\textsc{Submodular Knapsack-Constrained Maximization}} by a tuple $(I,f,w,B)$, in which $w=(w_i)_{i\in I}$ is the weight vector, and $B \in \R_{\geq 0}$ is the knapsack capacity. The \textbf{set of all instances} of \emph{\textsc{Submodular Knapsack-Constrained Maximization}} is denoted by $\mathcal{I}$.
For any $(I,f,w,B)\in \mathcal {I}$, we call a set $X\subseteq I$ an \textbf{optimal solution} if $w(X)\leq B$ and $f(X)$ attains the value given by~\eqref{e_prob1}.
\end{definition}

\hspace{-1.37pt}There are several well-known results for solving instances of \textsc{Submodular Knapsack-Constrained Maximization} approximately \parencite[see e.g.,][]{wolsey1982,sviridenko2004,tang2021,feldman2023}.  Among these, the algorithm proposed by~\textcite{sviridenko2004} stands out, since it provides an approximation factor of $1-\frac{1}{e} = 0.63\ldots$, which is the best achievable approximation factor unless $P=NP$, as shown by~\textcite{feige1998}.
However, achieving an approximation guarantee of $1-\frac{1}{e}$ may not be sufficient in many applications. Therefore, our focus is on the \emph{exact} solution of the \textsc{Submodular Knapsack-Constrained Maximization} problem.

\subsection{Results.}
We introduce a basic branch-and-bound algorithm for \textsc{Submodular Knapsack-\hspace{0pt}Constrained Maximization} that queries the objective function $f$ via a value oracle. To enhance its efficiency, we incorporate several acceleration techniques.

We evaluate the effectiveness of the proposed techniques on both artificial instances of three benchmark problems and instances based on real-world data. We compare them with the methods presented by~\textcite{sakaue2018}, which currently achieve the best performance reported in the literature~\parencite[see, e.g.,][]{sakaue2018}, and with a branch-and-cut algorithm implemented using Gurobi version 12.0.2 that solves a binary linear formulation of the submodular knapsack problem. The experimental results show that several of the proposed acceleration methods are highly effective, outperforming the algorithms of~\textcite{sakaue2018} and the branch-and-cut algorithm.

%%%%%%%%%%%%%%%%%%%%%%%%%%%%%%%%%%%%%%%%%%%%%%%%%%%%%%%%%%%%%%%%%%%%%%%%%%%%%%%%%%%%%%%%%%%%%%%%%%%%%%%%%%%%%%%%%%%

\subsection{Related work.}
Since the \textsc{Submodular Knapsack-Constrained Maximization} problem is NP-hard, previous research has focused primarily on developing approximation algorithms to tackle it. As mentioned above, \textcite{sviridenko2004} showed that combining partial enumeration with the classical greedy algorithm -- originally analyzed by~\textcite{wolsey1982} -- achieves a $1 - \frac{1}{e}$ approximation of the optimal solution value. Moreover,~\textcite{feige1998} proved that this is the best possible approximation ratio, unless $P = NP$.

In addition to these greedy-based approximate optimal solvers for \sloppy\textsc{Submodular Knapsack-Constrained Maximization},~\textcite{chen2015} introduced a best-first search algorithm for \textsc{Submodular Knapsack-Constrained Maximization}. This algorithm includes a parameter that controls its approximation guarantee, allowing it to serve as an exact solver for \textsc{Submodular Knapsack-Constrained Maximization}. Based on this work,~\textcite{sakaue2018} proposed two accelerated best-first search algorithms, called $\Umod$ and $\Udom$, both of which incorporate new termination conditions, with one additionally utilizing a novel heuristic function.

Several exact solvers have been proposed for the special case of \textsc{Submodular Knapsack-Constrained Maximization}, where all elements in the ground set have unit weights, which is the \textsc{Submodular Cardinality-Constrained Maximization} problem. \textcite{nemhauser1981} introduced two exact methods for solving this problem: a constraint generation algorithm that employs a binary integer linear program with exponentially many constraints and a branch-and-bound algorithm that utilizes linear programming relaxations.
An exact cutting-plane method for \textsc{Submodular Cardinality-Constrained Maximization}, implemented as a binary integer linear program with iterative constraint generation, was introduced by~\textcite{kawahara2009}.
\textcite{uematsu2020} enhanced~\citeauthor{nemhauser1981}'s constraint generation algorithm by integrating it into a branch-and-cut framework. More recently,~\textcite{csokas2024} improved the solver by~\textcite{uematsu2020} further by introducing alternative constraint generation heuristics and exploiting structural properties of specific problem instances.

\textcite{woydt2024} introduced a branch-and-bound algorithm for \textsc{Submodular Cardinality-Constrained Maximization}, which is a special case of \textsc{Submodular Knapsack-Constrained Maximization}, where all elements have unit weight, along with several acceleration techniques to enhance its efficiency. The branch-and-bound algorithm and the acceleration techniques for solving \textsc{Submodular Knapsack-Constrained Maximization} presented in this paper are inspired by their work. However, our methods differ significantly from those of~\textcite{woydt2024}, as we must account for a general knapsack constraint with non-unit weights. This places us in a more challenging setting, since~\textcite{woydt2024} did not have to consider weights at all, whereas in our case, the weights of the elements play a crucial role and require special attention.

%%%%%%%%%%%%%%%%%%%%%%%%%%%%%%%%%%%%%%%%%%%%%%%%%%%%%%%%%%%%%%%%%%%%%%%%%%%%%%%%%%%%%%%%%%%%%%%%%%%%%%%%%%%%%%%%%%%

\subsection{Outline.}
In Section~\ref{e_sec2}, we introduce a branch-and-bound algorithm for \textsc{Submodular Knapsack-Constrained Maximization}, which employs a pruning rule based on a linear relaxation of an integer program. 
We also introduce a variant of this branch-and-bound algorithm that directly solves the linear program using Gurobi and compare the performance of both algorithms (the detailed results are presented in Section~\ref{e_sec4}). Since our empirical tests show that the variant using the pruning rule based on a linear relaxation strictly outperforms the one using an exact solution to the integer program, we choose the first algorithm as our basic solver and propose several acceleration techniques for it, in Section~\ref{e_sec3}.

Section~\ref{e_sec4} presents our experimental results, where we evaluate the performance of the basic solver and the acceleration techniques on instances of three benchmark problems. We also compare the presented algorithms to the solvers $\Umod$ and $\Udom$ by~\textcite{sakaue2018} and to a branch-and-cut algorithm implemented using Gurobi that solves a binary linear reformulation of \textsc{Submodular Knapsack-Constrained Maximization}. 

For comparison, we generated artificial instances using two approaches. We first followed the method proposed by~\textcite{sakaue2018}. However, it turns out that those instances are easily solvable for any of the solvers we compare. Therefore, to provide a better comparison of the solvers, we also generated more challenging instances using our own method.
In addition, we evaluate all algorithms on instances based on real-world data. We discuss the construction of all instances in detail in Section~\ref{e_sec4}.

\section{A basic branch-and-bound algorithm.}\label{e_sec2}

In this section, we present a branch-and-bound algorithm for \textsc{Submodular Knapsack-Constrained Maximization}.
Given an instance $(I,f,w,B)\in \I$, it iteratively constructs a \textbf{search tree}, i.e., a directed graph $T=(V,A)$, in a depth-first manner.

\begin{definition}
Let $(I,f,w,B)\in \I$, and let $\mathcal{F} = \{S \subseteq I \colon w(S) \leq B\}$ denote the set of all feasible solutions to $(I,f,w,B)$. In the following, we assume that every $S\in \mathcal{F}$ is associated with a \textbf{candidate set} $C(S)\subseteq I\setminus S$.
A \textbf{basic} search tree $T=(V,A)$ of $(I,f,w,B)\in \I$ is a search tree that can be constructed iteratively according to the following procedure: We start with an empty search tree $T=(V,A)=(\emptyset,\emptyset)$. Then, we add the root node $\emptyset$ to $V$ and define the candidate set of the root node as $C(\emptyset)\coloneqq I$. For every node $S$ in the node set $V$, let the candidate set $C(S)=\{c_1,\dots,c_{\vert C(S)\vert}\}\subseteq I\setminus S$ be ordered by a fixed permutation. Then, any $S\cup\{c_i\}$ with $c_i\in C(S)$ and $w(S\cup\{c_i\})\leq B$ is added to $V$ and its candidate set $C(S\cup\{c_i\})$ is set as $C(S)\setminus \{c_1,\dots,c_i\}$. Furthermore, the arc $(S,S\cup\{c_i\})$ is added to the set of arcs $A$. We continue with this procedure until $V=\mathcal{F}$ and $A=\{(S,S\cup\{c\})\colon S,S\cup\{c\}\in\mathcal{F}, c\in C(S)\}$.
\end{definition}

The structure of a basic search tree is determined by the permutation of the set $C(S)$ for any node $S$ that is added to $V$.
In fact, this choice of the permutation plays a crucial role in the performance of our algorithms.

Basic search trees can be used to determine an optimal solution to any instance $(I,f,w,B)\in \I$ by evaluating the objective function at each node during the search tree's construction and continuously tracking the best solution value found so far.
It is straightforward that this search procedure returns the optimal solution value, as it evaluates all feasible solutions to the original problem. However, as any basic search tree might have exponentially many (in the number of elements) nodes, it is usually impractical.

As we go along, to avoid visiting each node of any basic search tree, we incorporate a simple pruning rule in the iterative construction of a basic search tree that attempts to determine whether any descendant of the current node can improve on the current best-known solution value. If no such improvement is possible, we skip constructing the children of the current node.

%%%%%%%%%%%%%%%%%%%%%%%%%%%%%%%%%%%%%%%%%%%%%%%%%%%%%%%%%%%%%%%%%%%%%%%%%%%%%%%%%%%%%%%%%%%%%%%%%%%%%%%%%%%%%%%%%%%

\subsection{A basic pruning rule.}
Consider an instance $(I,f,w,B)\in \I$ with a basic search tree $T=(V,A)$, and any $S\in V$. To decide whether a node $S$ can be pruned, we must determine whether any of its descendants can yield an improvement on the current best-known solution value, denoted as $s^*$. In particular, $S$ can be pruned if
\begin{align}
    \max_{X\subseteq C(S), w(X)\leq B- w(S)} f(S \cup X)\leq s^*.\label{e_in2}
\end{align}

However, this requires solving another submodular knapsack problem, which is NP-hard.
Nevertheless, we can prune $S$ if we can show that an upper bound on the left-hand side of Inequality~\eqref{e_in2} is smaller than or equal to $s^*$. To this end, we first define the marginal gain and relative marginal gain of an element $i\in I$, when added to a set $X\subseteq I$.

\begin{definition}
    Let $(I,f,w,B)\in\I$. For any $X,Y\subseteq I$, we define the \textbf{marginal gain of $Y$ with respect to $X$} as $f(Y\vert X)\coloneqq f(Y\cup X)-f(X)$. If $Y=\{i\}$ for some $i\in I$, we write $f(i\vert X)$ for short. Further, we define the \textbf{relative marginal gain} of an element $i\in I$ with respect to a set $X\subseteq I$ as $\frac{f(i\vert X)}{w_i}$.
\end{definition}

By exploiting the submodularity of $f$, we obtain the following upper bound on the left-hand side of Inequality~\eqref{e_in2}:

\begin{align}
    \max_{X\subseteq C(S), w(X)\leq B- w(S)} f(S \cup X) \leq  f(S) + \max_{X\subseteq C(S), w(X)\leq B-w(S)} \sum_{x\in X} f(x\vert S).\label{e_in3}
\end{align}

Notice that the second summand on the right-hand side of Inequality~\eqref{e_in3} is the optimal solution to the following knapsack problem:
\begin{align}
    \begin{aligned}
       &\max \sum_{x\in C(S)} p_xf(x\vert S)\\
    &\ \text{s.t. }\sum_{x\in C(S)}p_x w_x\leq B-w(S),\\
    &\ \phantom{\text{s.t. }}p_x\in\{0,1\}\text{ for }x\in C(S). 
    \end{aligned}
    \label{e_prob2}
\end{align}

The optimal solution value of this knapsack problem is upper bounded by the optimal solution value of the corresponding linear program, in which $p_x$ is relaxed to $[0,1]$, for all $x\in C(S)$. 
In the spirit of this linear program, we define the following:

\begin{definition}\label{e_def5}
Let  $(I,f,w,B)\in \I$ with a basic search tree $T=(V,A)$. For any $S\in V$ and $J\subseteq C(S)$, define
\begin{align*}
 b(S,J):=\max \left\{ \sum_{x\in J} p_xf(x\vert S) \colon \sum_{x\in J}p_x w_x\leq B-w(S),p_x\in [0,1] \text{ for }x\in J\right\}.
\end{align*}
\end{definition}

Notice that any value $b(S,J)$  can be computed efficiently by using a simple greedy algorithm.
This greedy algorithm initializes the fractional value $p_x$ of each element $x \in J$ as zero. It then processes the elements in non-increasing order of their relative marginal gain with respect to $S$, setting each $p_x$ to the largest possible value such that the knapsack constraint is still satisfied.

\begin{definition}\label{e_def6}
    Let $(I,f,w,B)\in\I$ with a basic search tree $T=(V,A)$. For any $S\in V$ and $J\subseteq C(S)$, the \textbf{fractional knapsack value} of $S$ and $J$ is defined by 
    \begin{align*}
         \fkh(S,J) \coloneqq f(S)+ b(S,J).
    \end{align*}
\end{definition}

Given an instance $(I,f,w,B)\in\I$, assume that we iteratively construct a basic search tree and proceed as described above to determine an optimal solution during its construction. Let $S$ be a node in the basic search tree, and let $s^*$ be the current best-known solution value.
If $\fkh(S,C(S)) \leq s^*$, then, by the definition of $\fkh$, it follows that $\max_{X\subseteq C(S), w(X)\leq B- w(S)} f(S \cup X)\leq s^*$, implying that $S$ can be pruned. 
Hence, by incorporating pruning via $\fkh$ into the search process, the resulting solution remains exact.

Recall that the structure of a basic search tree is determined by the permutation of the candidate set $C(S)$ of any $S\in V$. 
A natural choice for the ordering of the candidate sets is the order in which the greedy algorithm described above considers elements.

\begin{definition}
    Let $(I,f,w,B)\in\mathcal{I}$, $S\subseteq I$, and $C\subseteq I\setminus S$. 
    An order $\prec$ on $C$  is called a \textbf{greedy order with respect to $S$} if, for any $c,d \in C$ with $c \prec d$, the relative marginal gain of $c$ with respect to $S$ is greater than or equal to that of $d$, i.e.,
    \begin{align*}
       \frac{f(c\vert S)}{w_c}\geq \frac{f(d\vert S)}{w_d}.
    \end{align*}
    If the set $S$ is clear from context, we simply refer to $\prec$  as a greedy order.
\end{definition}
Note that several greedy orders may exist on a set due to tie-breaking. While it is possible to enforce uniqueness by defining a tie-breaking rule by a permutation of the set, such a restriction is unnecessary in our case; any greedy order will suffice.

In the following, we assume that the candidate set of each node is ordered according to a greedy order. Hence, the first leaf node reached when traversing the basic search tree in a depth-first manner corresponds to the solution produced by the greedy algorithm for \textsc{Submodular Knapsack-Constrained Maximization}, which tries to pack the elements based on their relative marginal gain, until each element is either packed or discarded, as described, e.g., by~\textcite{tang2021}.
Since~\textcite{feldman2023} showed that the objective value of the solution computed by this greedy algorithm achieves at least a $0.427$-approximation of the optimal solution, the same guarantee holds for the first leaf node reached.

The algorithm resulting from ordering each node’s candidate set according to a greedy order and incorporating a pruning rule based on $\fkh$ into the search process is presented in Algorithm~\ref{e_alg1}. 

\begin{algorithm}[H] 
\SetKwFunction{FSearch}{SearchNode}
\SetKwProg{Fn}{Function}{:}{}
\SetArgSty{textnormal}
\DontPrintSemicolon
\caption{}\label{e_alg1}
\KwIn{An instance $(I,f,w,B)\in\mathcal{I}$.}
\KwOut{Optimal objective value $s^*$}
\Fn{\FSearch{$S,C(S),s^*$}}{
    $s^* \gets \max\{s^*,f(S)\}$\;\label{e_alg1l2}
    \lIf{$C(S) = \emptyset$}{\KwRet{$s^*$}}
    \lIf{$\fkh(S,C(S))\leq s^*$ }{\KwRet{$s^*$}}\label{e_alg1l4}
    Let  $c_1,\ldots,c_k$  be the elements of $C(S)$ ordered by a greedy order.\label{e_alg1l5}\;
    \For{$i = 1,\dots,k$}{
        \mbox{\lIf{$w(S\cup \{c_i\})\leq B$}{$s^* \gets$ \FSearch{$S \cup \{c_i\},C(S)\setminus \{c_1,\dots,c_i\},s^*$}}}\label{e_alg1l7}  
    }
    \KwRet{$s^*$}
}
\vspace{0.2cm}
\textbf{Main:}\\
Initialize $s^* \gets -\infty$\;
\KwRet{\FSearch{$\emptyset,I,s^*$}}
\end{algorithm}

It is straightforward to demonstrate that Algorithm~\ref{e_alg1} terminates for any instance $(I,f,w,B)\in I$ and returns the objective value of an optimal solution to $(I,f,w,B)$.

\begin{theorem}
    Given an instance $(I,f,w,B)\in\I$, Algorithm~\ref{e_alg1} terminates and returns the objective value $s^*$ of an optimal solution to $(I,f,w,B)$.
\end{theorem}

\begin{proof}
Let $(I,f,w,B)\in \I$. First, consider Algorithm~\ref{e_alg1} with Line~\ref{e_alg1l4} excluded. Then Algorithm~\ref{e_alg1} constructs a complete basic search tree of $(I,f,w,B)$. Hence, it considers every node $S\in\mathcal{F}\coloneqq\{X\subseteq I\colon w(S)\leq B\}$. Since the current best-known solution value $s^*$ is updated in Line~\ref{e_alg1l2} of Algorithm~\ref{e_alg1} in each call to $\sn$, it follows directly that Algorithm~\ref{e_alg1} with Line~\ref{e_alg1l4} excluded returns the value of an optimal solution to $(I,f,w,B)$. 

Hence, it is sufficient to show that in Line~\ref{e_alg1l4} only nodes that cannot lead to an improvement over the current best solution are pruned, to demonstrate that Algorithm~\ref{e_alg1} returns the objective value of an optimal solution.
Let $S\in \mathcal{F}$ with candidate $C(S)$ be a node $\sn$ is called for in Algorithm~\ref{e_alg1}. Further, let $s^*$ be the current best-known solution value in Line~\ref{e_alg1l2} of Algorithm~\ref{e_alg1} in that call to $\sn$. Assume that the condition $\fkh(S,C(S))\leq s^*$ in Line~\ref{e_alg1l4} is satisfied. Then, by Definitions~\ref{e_def5} and~\ref{e_def6}, and submodularity of $f$, we have 
\begin{align*}
    s^*&\geq \fkh(S,C(S))=f(S)+b(S,C(S))\\
    &= f(S) + \max \{\sum_{c\in C(S)}p_cf(c\vert S)\colon \sum_{c\in C(S)}p_cw_c\leq B-w(S),p_c\in \left[0,1\right] \text{ for }c \in C(S)\} \\
    &\geq f(S) + \max\{\sum_{c\in X}f(c\vert S) \colon X\subseteq C(S), w(X)\leq B-w(S)\}\\
    &\geq \max \{f(S\cup X)\colon X\subseteq C(S), w(X)\leq B-w(S)\}.    
\end{align*}
Thus, no descendant of $S$ can achieve an improvement over the current best-known solution value $s^*$, and consequently Algorithm~\ref{e_alg1} returns the objective value of an optimal solution. Moreover, Algorithm~\ref{e_alg1}  terminates, since it calls $\sn$ for at most each node in a basic search tree, hence for every $S\in \mathcal{F}$ and $\mathcal{F}$ is finite as a set of subsets of the finite set $I$.
\end{proof}

A natural idea to accelerate Algorithm~\ref{e_alg1} is to improve the pruning rule in Line~\ref{e_alg1l4} by using a stronger upper bound on the left-hand side of Inequality~\eqref{e_in2}.
%%%%%%%%%%%%%%%%%%%%%%%%%%%%%%%%%%%%%%%%%%%%%%%%%%%%%%%%%%%%%%%%%%%%%%%%%%%%%%%%%%%%%%%%%%%%%%%%%%%%%%%%%%%%%%%%%%%

\subsubsection{Improving the pruning rule in Algorithm~\ref{e_alg1}.}

Let $(I,f,w,B)\in \I$ and $S$ be a node in a basic search tree with candidate set $C(S)$ for which $\sn(S,C(S),t^*)$ is called in Algorithm~\ref{e_alg1}. Further, let $s^*$ be the current best-known solution value set in Line~\ref{e_alg1l2} in that call to $\sn$. Then Algorithm~\ref{e_alg1} uses a pruning rule to determine whether any of the descendants of $S$ can achieve an improvement over $s^*$. If such an improvement is impossible, Algorithm~\ref{e_alg1} avoids considering the descendants of $S$. 

Recall that the node $S$ can be pruned if Inequality~\eqref{e_in2} is valid, implying that $S$ can be pruned if any upper bound on the left-hand side of Inequality~\eqref{e_in2} is smaller than or equal to $s^*$.
The pruning rule used in Algorithm~\ref{e_alg1} checks whether $\fkh(S,C(S))\leq s^*$, where the fractional knapsack value $\fkh(S,C(S))$ is an upper bound on the left-hand side of Inequality~\eqref{e_in2}, which is based on the linear relaxation of Problem~\eqref{e_prob2}.
To improve the pruning rule, it is a natural idea to tighten the upper bound on the left-hand side of Inequality~\eqref{e_in2}, that is, to obtain a tighter (i.e., smaller) upper bound than $\fkh(S,C(S))$. To achieve such a tighter upper bound, we now solve Problem~\eqref{e_prob2} exactly rather than its linear relaxation.
In this spirit, we define the following: 
\begin{definition}
    Let $(I,f,w,B)\in \I$ with a basic search tree $T=(V,A)$. For any $S\in V$ and $J\subseteq C(S)$, the knapsack value of $S$ and $J$ is defined by
    \begin{align*}
        \kv(S,J)\coloneqq f(S) +\max \left\{\sum_{x\in J} p_x f(x\vert S)\colon \sum_{x\in J}p_x w_x\leq B-w(S),p_x\in \{0,1\} \text{ for }x\in J\right\}.
    \end{align*}
\end{definition}
By replacing the condition $\fkh(S, C(S)) \leq s^*$ in Line~\ref{e_alg1l4} of Algorithm~\ref{e_alg1} with $\kv(S, C(S)) \leq s^*$, we obtain a variant of the algorithm, which we refer to as Algorithm~\ref{e_alg1}$^+$.

Since $\kv(S,C(S))\leq \fkh(S,C(S))$ by definition, Algorithm~\ref{e_alg1}$^+$ prunes at least as many nodes as Algorithm~\ref{e_alg1}, and possibly more.
However, this improved pruning comes at the cost of solving the NP-hard Problem~\eqref{e_prob2} in every call to $\sn$ in Algorithm~\ref{e_alg1}. 
We evaluate the performance of both algorithms in Section~\ref{e_sec4} on instances of three benchmark problems. To compute $\kv(S, C(S))$ in each call to $\sn$ in Algorithm~\ref{e_alg1}$^+$, we solve Problem~\eqref{e_prob2} using Gurobi version 12.0.2. 
We describe our experimental setup and present the results of our experiments in detail in Section~\ref{e_sec4}. However, we would like to preemptively highlight here that Algorithm~\ref{e_alg1} consistently outperformed Algorithm~\ref{e_alg1}$^+$ in all of our tests. Hence, the increase in the number of pruned nodes achieved by using $\kv$ instead of $\fkh$ does not justify the effort of solving the NP-hard Problem~\eqref{e_prob2}.
Consequently, we use Algorithm~\ref{e_alg1} as the basic solver in the remainder of this paper.

%%%%%%%%%%%%%%%%%%%%%%%%%%%%%%%%%%%%%%%%%%%%%%%%%%%%%%%%%%%%%%%%%%%%%%%%%%%%%%%%%%%%%%%%%%%%%%%%%%%%%%%%%%%%%%%%%%%%%%%%%%%%%%%%%%%%%%%%%%%%%%%%%%%%%%%%%%%%%%%%%%%%%%%%%%%%%%%%%%%%%%%%%%%%%%%%%%%%%%%%%%%%%%%%%%%%%%%%%%%%%%%%%%%%%%

\section{Acceleration techniques.}\label{e_sec3}

In this section, we present three acceleration techniques for Algorithm~\ref{e_alg1}: \emph{Lazy Evaluations}, \emph{Early Pruning}, and \emph{Candidate Reduction}. All three aim to reduce computational effort by limiting either expensive objective function evaluations or unnecessary branching during the search.

Lazy Evaluations improve the efficient computation of $\fkh$ by selectively avoiding evaluations of the objective function, thereby accelerating the pruning rule in Algorithm~\ref{e_alg1}.
Similarly, Early Pruning seeks to avoid evaluations of the objective function by performing pruning as early as possible.
If a node cannot be pruned, Candidate Reduction reduces the number of considered descendants of this node by preventing the generation of branches that cannot improve the so far best-known solution value.

%%%%%%%%%%%%%%%%%%%%%%%%%%%%%%%%%%%%%%%%%%%%%%%%%%%%%%%%%%%%%%%%%%%%%%%%%%%%%%%%%%%%%%%%%%%%%%%%%%%%%%%%%%%%%%%%%%%

\subsection{Lazy Evaluations.}
Given an instance $(I,f,w,B)\in \I$, Algorithm~\ref{e_alg1} uses the fractional knapsack value to determine whether a node $S$ in the current call of $\sn$  can be pruned.
Recall that to compute $\fkh(S, C(S))$, for any node $S$ with candidate set $C(S)$, we need to solve
\begin{align*}
    b(S,C(S))=\max\!\left\{\sum_{c\in C(S)} p_c f(c\vert S) \colon\! \sum_{c\in C(S)}p_c w(c)\leq B-w(S),p_c\in [0,1] \text{ for }c\in C(S)\right\}\!.
\end{align*}

As described in Section~\ref{e_sec2}, this can be done efficiently using a greedy algorithm that requires comparing $\frac{f(c\vert S)}{w_c}$ for all $c\in C(S)$. 
To determine $b(S,C(S))$, it is often sufficient to compute only the relative marginal gains of a specific subset of $C(S)$. Since the greedy algorithm processes the elements in $C(S)$ in non-increasing order of their relative marginal gains, and because the total weight of all elements $c\in C(S)$ for which the greedy algorithm sets $p_c>0$ often reaches $B-w(S)$ before all elements have been processed, it is sufficient to compute only the (relative) marginal gains of these elements to determine $b(S,C(S))$.
However, since we cannot know in advance for which elements $c\in C(S)$ the greedy algorithm sets $p_c>0$ and for which $p_c=0$ remains unchanged, it is not possible to simply omit the computation of marginal gains for any elements.
Instead, we apply the concept of ``Lazy Evaluations'', introduced by~\textcite{minoux2005}, by replacing the marginal gain $f(c\vert S)$ of an element $c\in C(S)$, for which we assume it is unlikely that the greedy algorithm sets $p_c>0$, with a natural upper bound such as $f(c\vert S^P)$, the marginal gain of $c$ with respect to the parent node $S^P$. Replacing $f(c\vert S)$ with an upper bound when applying the greedy algorithm ensures that, in case the greedy algorithm sets $p_c>0$, although we had considered this unlikely, it still computes an upper bound on $b(S,C(S))$ and thus the pruning rule in Algorithm~\ref{e_alg1} is applied using an upper bound on $\fkh(S,C(S))$. As a result, Algorithm~\ref{e_alg1} continues to prune only nodes that cannot achieve an improvement over the best solution value found so far.
To specify for which elements we omit the exact computation of $f(c\vert S)$, we need the following two definitions:

\begin{definition}
    Let $(I,f,w,B)\in\I$ with a basic search tree $T=(V,A)$. For any $\emptyset\neq S\in V$, we call a function $\pi^S\colon C(S) \to \{0,1\}$ a \textbf{decision function}.
\end{definition}

\begin{definition}
    Let $(I,f,w,B)\in\I$ with a basic search tree $T=(V,A)$, and for any $\emptyset\neq S\in V$, let  $\pi^S\colon C(S) \to \{0,1\}$ be a decision function. Then, for any $S\in V$, the function $u^T_S\colon C(S)\to \R_{\geq 0}$, 
    \begin{align*}
        u^{T}_S(c)\coloneqq\begin{cases}
            f(c) &\text{if } S = \emptyset, \\
            \pi^S(c) f(c\vert S)+(1-\pi^S(c))u^T_{S^P}(c) &\text{if } S\neq \emptyset,
        \end{cases}
    \end{align*}
is called a \textbf{lazy evaluation scheme in $S$ on $T$}, and $\frac{u^T_S(c)}{w_c}$ is called the \textbf{lazy relative marginal gain} of $c$ with respect to $S$ regarding $T$. If the context clearly indicates which search tree $T$ we refer to, we use $u_S$ to represent $u_S^T$. 
\end{definition}

Given an instance $(I, f, w, B)$ with a basic search tree and a node $S \neq \emptyset$ with a decision function $\pi^S$, the computation of the marginal gain of an element $c\in C(S)$ is avoided if $\pi^S(c) = 0$.
In this case, a corresponding lazy evaluation scheme inherits the upper bound $u^T_{S^P}(c)$ on the marginal gain of $c$ from the parent node $S^P$ to the current node $S$.
To formally incorporate the concept of lazy evaluation schemes in Algorithm~\ref{e_alg1}, we define the analogous to the fractional knapsack value with respect to a given lazy evaluation scheme.
\begin{definition}
Let $(I,f,w,B)\in\I$ with a basic search tree $T=(V,A)$.
For any $S\in V$, and any $J\subseteq C(S)$, we define the \textbf{lazy fractional knapsack value} with respect to the lazy evaluation scheme $u^T_S$ in $S$ on $T$ by
\begin{align*}
\lfkh(S, J,u^T_S):=f(S)+\max \left\{ \sum_{c\in J}p_c u^T_S(c) \colon\!\sum_{c\in J}p_c w(c)\leq B-w(S), p_c\in [0,1]\text{ for } c\in J\right\}\!.  
\end{align*}
\end{definition}

Algorithm~\ref{e_alg1} is exact, and for any lazy evaluation scheme $u^T_S$ in $S$ on a basic search tree $T$, we have $\lfkh(S,C(S),u^T_S)\geq \fkh(S,C(S))$. Therefore, Algorithm~\ref{e_alg1} remains exact when replacing $\fkh(S,C(S))$ with $\lfkh(S,C(S),u^T_S)$ in Line~\ref{e_alg1l4}.

Furthermore, we modify Algorithm~\ref{e_alg1} by ordering the candidate set $C(S)$ in each call of $\sn(S,C(S),s^*)$ according to the ``lazy'' equivalent of a greedy order.

\begin{definition}
    Let $(I,f,w,B)\in\I$ with a basic search tree $T=(V,A)$ and $S\in V$. For any lazy evaluation scheme $u^T_S$, we call an order $\prec$ on $C(S)$ a \textbf{$u^T_S$-greedy order} if for any $c,d\in C(S)$ with $c\prec d$, we have
    \begin{align*}
        \frac{u^T_S(c)}{w_c}\geq \frac{u^T_S(d)}{w_d}.
    \end{align*}
\end{definition}
Notice that in general, for any node $S$ in a basic search tree $T$, a greedy order on the candidate set $C(S)$ does not coincide with a $u^{T}_S$-greedy order on $C(S)$. 

In summary, the modification of Algorithm~\ref{e_alg1} to incorporate Lazy Evaluations is achieved by replacing Lines~\ref{e_alg1l4} and~\ref{e_alg1l5} with

{\LinesNotNumbered
\DontPrintSemicolon
\algorithmstyle{tworuled}
\begin{algorithm}[H]
\nlset{4$^{\text{LE}}$}  \lIf{$\lfkh(S,C(S),u_S)\leq s^*$ }{\KwRet{$s^*$}}
\nlset{5$^{\text{LE}}$} Let  $c_1,\ldots,c_k$ be the elements of $C(S)$ ordered by a $u_S$-greedy order.
\end{algorithm}}

The lazy evaluation scheme in a node $S\neq \emptyset$ of a basic search tree $T$ depends primarily on the chosen decision function. 
In the following, we present two specific decision functions, which we use in each call to $\sn$ when integrating Lazy Evaluations into Algorithm~\ref{e_alg1}.
The \emph{first} decision function is based on the simple idea that, given a node $S$, it is unlikely that the greedy Algorithm sets $p_c>0$ for an element $c\in C(S)$, or, respectively, that $c\in C(S)$ contributes to the lazy fractional knapsack value in $S$, if it has not contributed to the lazy fractional knapsack value in the parent node $S^P$.
Accordingly, the lazy evaluation scheme using the \textbf{greedy decision rule}, defined below, computes the exact marginal gain $f(c\vert S)$ only for those $c\in C(S)$ that contribute to the lazy fractional knapsack value in $S^P$, while all other elements inherit an upper bound on their marginal gain from the parent node $S^P$.

\begin{definition}
   Let $(I,f,w,B)\in\I$ with a basic search tree $T=(V,A)$. 
  For any node $\emptyset\neq S\in V$, let $u^T_{S^P}$ be a lazy evaluation scheme in $S^P$ on $T$. Further, let $C(S^P)=\{c_1^P,\dots,c_{\vert C(S^P)\vert}^P\}$ be the candidate set of $S^P$ ordered according to a $u^T_{S^P}$-greedy order $\prec_g$ and $j^*=\min\{j\in\{1,\dots, \vert C(S^P)\vert \}\colon w(\{c^P_1,\dots,c^P_j\})\geq B-w(S)\}$. Then, the \textbf{greedy decision rule} $\pi^S\colon C(S)\to \R_{\geq 0}$ is defined by 
  \begin{align*}
        \pi^S(c)\coloneqq\begin{cases}
            1 &\text{if } c \preceq_g c^P_{j^*} ,\\
            0 &\text{otherwise}.
        \end{cases}
    \end{align*}  
\end{definition}
The \emph{second} decision function is based on the following idea: In a call to $\sn(S,C(S),t^*)$, the exact marginal gain of an element $c\in C(S)$ should only be computed if $c$ offers the possibility that the lazy fractional knapsack value in $S$ exceeds the current best-known solution value $s^*$, as set in Line~\ref{e_alg1l2} of Algorithm~\ref{e_alg1}, which means that the node $S$ cannot be pruned. Therefore, we want the exact marginal gain of an element $c\in C(S)$ to be computed only if the lazy relative marginal gain of $c$ with respect to $S^P$ is greater than or equal to the average relative marginal gain required for any element in $C(S)$ to potentially improve the current best-known solution value $s^*$ in any descendant of $S$.

\begin{definition}
   Let $(I,f,w,B)\in\I$ with a basic search tree $T=(V,A)$ and let $k\colon V\to \R_{\geq 0}$ be a function with $k\geq f$. 
  For any node $\emptyset\neq S\in V$, let $u^T_{S^P}$ be a lazy evaluation scheme in $S^P$ on $T$. Then, the \textbf{average decision rule} $\pi^S\colon C(S)\to \R_{\geq 0}$ is defined by 
  \begin{align*}
        \pi^S(c)\coloneqq\begin{cases}
            1 &\text{if } \frac{u^T_{S^P}(c)}{w_c} \geq \frac{k(S)-f(S)}{B-w(S)} ,\\
            0 &\text{otherwise}.
        \end{cases}
    \end{align*}  
\end{definition}

To ensure that a lazy update scheme using the average decision rule computes the marginal gain of an element only in the case described above, we apply the average decision rule in each call to $\sn(S,C(S),t^*)$ for a node $S\neq\emptyset$ with $k(S)=s^*$, where $s^*$ is the current best-known solution value from Line~\ref{e_alg1l2} of Algorithm~\ref{e_alg1} in that call.

In the following, we use $\LEg$ and $\LE$ to refer to Algorithm~\ref{e_alg1} with a lazy evaluation scheme based on the greedy decision rule or the average decision rule, respectively, incorporated.

%%%%%%%%%%%%%%%%%%%%%%%%%%%%%%%%%%%%%%%%%%%%%%%%%%%%%%%%%%%%%%%%%%%%%%%%%%%%%%%%%%%%%%%%%%%%%%%%%%%%%%%%%%%%%%%%%%%%%%%%%%%%%%%%%%%%%%%%%%%%%%%%%%%%%%%%%%%%%%%%%%%%%%%%%%%%%%%%%%%%%%%%%%%%%%%%%%%%%%%%%%%%%%%%%%%%%%%%%%%%%%%%%%%%%%%%%%%%%%%%%%%%%%%%%%%%%%%%%%%%%%%%%%%%%%%%%%%%%%%%%%%%%%%%%%%%%%%%%%%%%%%%%%%%%%%%%%%%%%%%%%%%%%%%%%%%%%%%%%%

\subsection{Early Pruning.}

Given an instance $(I,f,w,B)\in \I$ with a basic search tree $T=(V,E)$, the idea behind Lazy Evaluations is to accelerate the pruning rule in Line~\ref{e_alg1l4} of Algorithm~\ref{e_alg1} by avoiding the computation of each marginal gain $f(c\vert S)$, with $\emptyset\neq S\in V$ and $c\in C(S)$. Instead, it is substituted by an upper bound.
Similarly, Early Pruning aims to reduce the number of computations of (relative) marginal gains.
However, unlike Lazy Evaluations, Early Pruning does not aim to replace the marginal gains with an upper bound. Instead, as the name Early Pruning suggests, it aims to perform pruning earlier than in Algorithm~\ref{e_alg1}, specifically while ordering the elements in the candidate set $C(S)$ of a node $S \neq \emptyset$ by a greedy order.

In any call to $\sn(S,C(S),s^*)$ in Algorithm~\ref{e_alg1}, we consider the elements $c_1,\dots,c_k$ of $C(S)$  sorted according to a greedy order both to efficiently compute $\fkh(S,C(S))$ and to determine the order in which Algorithm~\ref{e_alg1} explores the children of $S$. In practice, we order the candidate set $C(S)$ iteratively by considering each element $c_i\in C(S)$, computing its relative marginal gain $\frac{f(c_i\vert S)}{w_{c_i}}$, and inserting it into the correct position within the current partial greedy order of the previously considered elements $\{c_1,\dots,c_{i-1}\}$.

To avoid computing relative marginal gains $\frac{f(c\vert S)}{w_c}$ for some $c\in C(S)$, we incorporate a pruning rule into the iterative calculation of the relative marginal gains, based on the relative marginal gains computed in earlier iterations and the relative marginal gains $\frac{f(c\vert S^P)}{w_c}$ of all $c\in C(S)$ with respect to the parent node $S^P$.

\begin{theorem}\label{e_th2}
    Let $(I,f,w,B)\in \I$ with a basic search tree $T=(V,A)$ and let $k\colon V\to \R_{\geq 0}$ be a function with $k\geq f$. Let $\emptyset\neq S\in V$ and let $U=\{c_1,\dots,c_i\}$ with $i\leq \vert C(S)\vert$ be a subset of $C(S)$ that is ordered according to a greedy order and satisfies $w(U)\geq B-w(S)$ and $\fkh(S,U)\leq k(S)$.
    Further, let $j^*= \min \{j\in \{1,\dots,i\}\colon w(\{c_1,\dots,c_j\})\geq B-w(S)\}$ with $\frac{f(c_{j^*}\vert S)}{w_{c_{j^*}}}\geq \max\{\frac{f(c\vert S^P)}{w_c}\colon c\in C(S)\setminus U\}$.
    Then, for any $X\subseteq C(S)$ with $w(X)\leq B-w(S)$, we have 
    \begin{align*}
        f(S\cup X)\leq k(S).
    \end{align*}
\end{theorem}

\begin{proof}
    For each $c\in C(S)\setminus U$, we have
    \begin{align*}
        \frac{f(c_{j^*}\vert S)}{w_{c_{j^*}}} \geq \frac{f(c\vert S^p)}{w_c} \geq \frac{f(c\vert S)}{w_c},
    \end{align*}
    since $f$ is submodular.
    Thus, when ordering $C(S)$ according to a greedy order, all $c\in C(S)\setminus U$ are placed after $c_{j^*}$. Since $w(U)\geq B-w(S)$ and $j^*= \min \{j\in \{1,\dots,i\}\colon w(\{c_1,\dots,c_j\})\geq B-w(S)\}$, this directly implies that the greedy algorithm described in Section~\ref{e_sec2}, returns the same solution for the two problems 
    \begin{align*}
        \max\{\sum_{c\in U}p_cf(c\vert S)\colon \sum_{c\in U}p_cw_c\leq B-w(S), p_c\in \left[0,1\right] \text{ for } c\in U\}
    \end{align*}
    and
    \begin{align*}
        \max\{\sum_{c\in C(S)}p_cf(c\vert S)\colon \sum_{c\in C(S)}p_c w_c\leq B-w(S), p_c\in \left[0,1\right] \text{ for } c\in C(S)\}.
    \end{align*}
    Hence, $\fkh(S,U)=\fkh(S,C(S))$. 
    
    Assume that there exists a set $X\subseteq C(S)$ with $w(X)\leq B-w(S)$ and $f(S\cup X) > k(S)$. Then, we have
    \begin{align*}
        f(S\cup X) &> k(S) \geq \fkh(S,U) = \fkh(S,C(S)) = f(S) + b(S,C(S)) \\
        &\geq \max_{Y\subseteq C(S), w(Y)\leq B-w(S)} f(S\cup Y),
    \end{align*}
    which yields a contradiction and thereby proves the claim.
\end{proof}

Theorem~\ref{e_th2} provides a straightforward rule for pruning a node $S\neq\emptyset$ during the iterative ordering of the candidate set $C(S)$. To incorporate this pruning rule into Algorithm~\ref{e_alg1}, we set $k(S)=s^*$ for each node $S\neq \emptyset$ on which $\sn$ is called, with $s^*$ being the current best-known solution value in that call, as set in Line~\ref{e_alg1l2} of Algorithm~\ref{e_alg1}.
Accordingly, we define the following:

\begin{definition}
    Let $(I,f,w,B)\in \I$ with a basic search tree $T=(V,A)$. Let $\emptyset \neq S\in V$ be a node for which $\sn(S,C(S),t^*)$ is called in Algorithm~\ref{e_alg1}. Further, let $U\subseteq C(S)$ and $s^*$ be the current best-known solution value in that call to $\sn$, as set in Line~\ref{e_alg1l2} of Algorithm~\ref{e_alg1}. If $S$ and $U$ satisfy the conditions in Theorem~\ref{e_th2} with $k(S)=s^*$, we say that $S$ and $U$ satisfy the \textbf{early pruning conditions}. 
    
    If $S$ and $U$ only satisfy $w(U)\geq B-w(S)$ and $\frac{f(c_{j^*}\vert S)}{w_{c_{j^*}}}\geq \max\{\frac{f(c\vert S^P)}{w_c}\colon c\in C(S)\setminus U\}$, where $j^*$ is defined as in Theorem~\ref{e_th2}, but not $\fkh(S,U)\leq s^*$, we say that $S$ and $U$ satisfy the \textbf{early no-pruning conditions}.
\end{definition}

Consider an instance $(I,f,w,B)\in \I$ with basic search tree $T=(V,A)$. Let $\emptyset \neq S\in V$ be a node and let $\{c_1,\dots,c_i\}\subseteq C(S)$. If $S$ and $\{c_1,\dots,c_i\}$ satisfy the early pruning conditions, it follows by Theorem~\ref{e_th2}, that we can prune node $S$ without calculating the marginal gains $f(c\vert S)$ for $c\in C(S)\setminus \{c_1,\dots,c_i\}$. 

If $S$ and $\{c_1,\dots,c_i\}$ satisfy the early no-pruning conditions, it follows directly that $\fkh(S,\{c_1,\dots,c_i\})=\fkh(S,C(S))>s^*$. 
Therefore, $S$ cannot be pruned, and the early pruning conditions cannot be satisfied by any subset $X\subseteq C(S)$ with $\{c_1,\dots,c_i\}\subseteq X$; thus, there is no need to check them for any such a set $X\subseteq C(S)$.

To incorporate a pruning rule based on the early pruning conditions (and the early no-pruning conditions) into Algorithm~\ref{e_alg1}, we replace Lines~\ref{e_alg1l4} and~\ref{e_alg1l5} in each call to $\sn(S,C(S),t^*)$, with $S \neq \emptyset$, by the following:

{\LinesNotNumbered
\DontPrintSemicolon
\algorithmstyle{tworuled}
\SetArgSty{textnormal}
\begin{algorithm}[H]
\nlset{4$^{\text{EP}}$} Let $c_1,\dots,c_k$ be the elements of $C(S)$ ordered by a greedy order w.r.t. $S^P$\label{e_4EP}.\;
\nlset{5$^{\text{EP}}$}  \For{$i=1,\dots,k$\label{e_8EP}}{
\nlset{6$^{\text{EP}}$}  Compute $\frac{f(c_i\vert S)}{w_i}$\;
\nlset{7$^{\text{EP}}$}  Place $c_i$ into a correct position within the elements $c_1,                                                    \dots,c_{i-1}$ ordered by a greedy order w.r.t. $S$. (e.g. using a heap)\;
\nlset{8$^{\text{EP}}$}  \lIf{$S$ and $\{c_1,\dots,c_{i}\}$ satisfy the early pruning conditions}                                            {\KwRet{$s^*$}}     
\nlset{9$^{\text{EP}}$}   \lElseIf{$S$ and $\{c_1,\dots,c_{i}\}$ satisfy the early no-pruning conditions}{\textbf{break}}}
\nlset{10$^{\text{EP}}$}  Let $c_1,\dots,c_k$ be the elements of $C(S)$ ordered by a greedy order w.r.t. $S$\label{e_10EP}.
\end{algorithm}}

Notice that in Line~\ref{e_4EP}, the elements of $C(S)$ are ordered according to a greedy order with respect to $S^P$. This ordering ensures that, in each iteration $i \in \{1, \dots, k\}$ of the for-loop in Line~\ref{e_8EP}, we can check whether the conditions for early pruning or early no-pruning are satisfied without explicitly computing the maximum relative marginal gain of adding an element from $C(S) \setminus \{c_1, \dots, c_i\}$ to $S^P$, since this maximum is always attained by the next element $c_{i+1}$ in the order. 
Moreover, the elements in $C(S)$ are already ordered by a greedy order with respect to $S^P$ during the previous call to $\sn$ for $S^P$, so no additional sorting is required in Line~\ref{e_4EP}.

Further, observe that if the for-loop in Line~\ref{e_8EP} runs to completion, the elements in $C(S)$ are already ordered by a greedy order with respect to $S$, and in Line~\ref{e_10EP} no further computations are needed. 

%%%%%%%%%%%%%%%%%%%%%%%%%%%%%%%%%%%%%%%%%%%%%%%%%%%%%%%%%%%%%%%%%%%%%%%%%%%%%%%%%%%%%%%%%%%%%%%%%%%%%%%%%%%%%%%%%%%%%%%%%%%%%%%%%%%%%%%%%%%%%%%%%%%%%%%%%%%%%%%%%%%%%%%%%%%%%%%%%%%%%%%%%%%%%%%%%%%%%%%%%%%%%%%%%%%%%%%%%%%%%%%%%%%%%%%%%%%%%%%%%%%%%%%%%%%%%%%%%%%%%%%%%%%%%%%%%%%%%%%%%%%%%%%%%%%%%%%%%%%%%%%%%%%%%%%%%%%%%%%%%%%%%%%%%%%%%%%%%%%

\subsection{Candidate Reduction.}
Given an instance $(I,f,w,B)\in \I$, and a node $S$ in a basic search tree that cannot be pruned in Algorithm~\ref{e_alg1}, we aim to avoid considering all children of $S$. We only wish to explore children (and their descendants) that may lead to an improvement over the currently best-known solution value. 
To this end, we assume in the following that each node $S$ in a basic search tree is associated not only with a candidate set $C(S)$, but also with a subset of $C(S)$ called \textbf{reduction set}.

\begin{definition}
    Let $(I,f,w,B)\in \I$ with a basic search tree $T=(V,A)$, and let $k\colon V\to \R_{\geq 0}$ be a function with $k\geq f$. Then, the \textbf{reduction set} of $S$ with respect to $k$ is defined as 
    \begin{align*}
        R^k(S)\coloneqq \left\{c\in C(S)\colon f(c\vert S)+\fkh(S,C(S)\setminus\{c\})\leq k(S)\right\}.
    \end{align*}
\end{definition}

An element $r\in C(S)$ is contained in $R^k(S)$ if the marginal gain $f(r\vert S)$ plus the fractional knapsack value of $S$ and $C(S)\setminus \{r\}$ is less than or equal to $k(S)$. 

We demonstrate that for any $r\in R^k(S)$, there exists no set $X\subseteq C(S)\setminus\{r\}$ such that $ S\cup \{r\}\cup X$ is a feasible solution with objective value greater than $k(S)$.

\begin{theorem}\label{e_th3}
    Let $(I,f,w,B)\in \I$ with a basic search tree $T=(V,A)$ and let $k\colon V\to \R_{\geq 0}$ be a function with $k\geq f$. Let $S\in V$ and $r\in R^k(S)$.
    Then, for any $X\subseteq C(S)\setminus\{r\}$ with $w(X)\leq B-w(S\cup\{r\})$, we have 
    \begin{align*}
         f(S\cup\{r\}\cup X)\leq k(S).
    \end{align*}
\end{theorem}
\begin{proof}
    Assume that there exists a set $X\subseteq C(S)\setminus\{r\}$ with $w(X)\leq B-w(S\cup\{r\})$ and $f(S\cup\{r\}\cup X)> k(S)$.
    Then, we have
    \begin{align*}
      f(S\cup X)   \underset{\text{submodularity}}{\geq}& f(S\cup\{r\}\cup X)- f(S\cup \{r\})\,+\;f(S)  \underset{\text{assumption}}{>} k(S) - f(r\vert S)\\ \underset{r\in R^k(S)}{\geq} &\fkh(S,C(S)\setminus\{r\}),
    \end{align*} 
    which yields a contradiction by the definition of $\fkh$.
\end{proof}

Given an instance \((I, f, w, B)\), suppose that for every \(S \in V\), the value \(k(S)\) is a \emph{lower} bound on the optimal objective value of \((I, f, w, B)\). Then, by Theorem~\ref{e_th3}, it follows that for any \(S \in V\) and any \(r \in R^k(S)\), the subtree rooted at \(S \cup \{r\}\) can be pruned, since the lower bound \(k(S)\) on the optimal objective value cannot be exceeded by the objective value of \(S \cup \{r\}\) or any of its descendants.
To exclude as many children of $S$ and their descendants from consideration as possible, we aim to choose $k(S)$ to be a lower bound on the optimal objective value that is as large as possible. 
Since $s^*$ in Line~\ref{e_alg1l2} of Algorithm~\ref{e_alg1} represents the best-known objective value in each call of $\sn$, and is itself a lower bound on the optimal objective value, a natural choice is to set $k(S)=s^*$ for every $S$ where $\sn$ is invoked. We refer to this approach as Candidate Reduction, and for the reduction set of a node $S$ resulting from this particular choice of $k$, we write $R(S)$. 

To incorporate Candidate Reduction into Algorithm~\ref{e_alg1}, we replace in each call to $\sn(S,C(S),t^*)$, with $S \neq \emptyset$, Lines~\ref{e_alg1l4} with the following two lines 

{\LinesNotNumbered
\DontPrintSemicolon
\algorithmstyle{tworuled}
\begin{algorithm}[H]
\nlset{4$^{\text{CR}}$} \lIf{$\fkh(S,C(S)\setminus R(S^P))\leq s^*$ }{\KwRet{$s^*$}}
\nlset{5$^{\text{CR}}$} \mbox{Set $R(S)\!\leftarrow\!(R(S^P)\cap C(S))\cup\{c\in C(S)\setminus R(S^P)\colon\!f(c\vert S)\!+\!\fkh(S,C(S)\setminus\{c\})\leq s^* \}$}\label{e_5CR}
\end{algorithm}}

\noindent and Line~\ref{e_alg1l7} with

{\LinesNotNumbered
\DontPrintSemicolon
\algorithmstyle{tworuled}
\begin{algorithm}[H]
\nlset{7$^{\text{CR}}$} \If{$w(S\cup \{c_i\})\leq B$ and $c_i\not\in R(S)$}{\nlset{8$^{\text{CR}}$}$s^* \gets$ \FSearch{$S \cup \{c_i\},C(S)\setminus \{c_1,\dots,c_i\},s^*$}}
\end{algorithm}}
\noindent In the initial call $\sn(\emptyset,C(\emptyset),t^*)$, we keep Line~\ref{e_alg1l4} of Algorithm~\ref{e_alg1} unchanged and introduce the other modifications as described for the other calls to $\sn$, except that we set $R(S) \leftarrow \{c\in C(S)\colon f(c\vert S) + \fkh(S,C(S)\setminus\{c\})\leq s^* \}$ in Line~\ref{e_5CR}. We refer to the algorithm resulting from these modifications as $\CR$.

Notice that reduction sets in $\CR$ are partly inherited by descendant nodes. Specifically, let $S$ be a node for which $\sn$ is called in $\CR$. Then, any element $r\in R(S)$ is also contained in the reduction set $R(S^\prime)$ of any child $S^\prime$ of $S$ if $r\in C(S^\prime)$.  
Hence, when calling $\sn$ for a node $S\neq \emptyset$ in $\CR$, it is sufficient to check if $\fkh(S, C(S)\setminus R(S^P))\leq s^*$ to prune $S$. If a node $S\neq\emptyset$ can not be pruned, we determine the reduction set 
$R(S)$ by taking the union of all elements $C(S)\cap R(S^P)$, which we inherit from the reduction set of the parent node, with all elements from $C(S)\setminus R(S^P)$ that satisfy the condition $f(c\vert S)+\fkh(S,C(S)\setminus\{c\})\leq s^*$.

%%%%%%%%%%%%%%%%%%%%%%%%%%%%%%%%%%%%%%%%%%%%%%%%%%%%%%%%%%%%%%%%%%%%%%%%%%%%%%%%%%%%%%%%%%%%%%%%%%%%%%%%%%%%%%%%%%%%%%%%%%%%%%%%%%%%%%%%%%%%%%%%%%%%%%%%%%%%%%%%%%%%%%%%%%%%%%%%%%%%%%%%%%%%%%%%%%%%%%%%%%%%%%%%%%%%%%%%%%%%%%%%%%%%%%

\section{Experiments.}\label{e_sec4}
In this section, we present the results of our computational experiments. We evaluate the performance of our basic algorithm (Algorithm~\ref{e_alg1}) and its variation Algorithm~\ref{e_alg1}$^+$, as well as the accelerated variants of Algorithm~\ref{e_alg1}, which incorporate Lazy Evaluations with the greedy decision rule ($\LEg$) and the average decision rule $(\LE)$, Early Pruning ($\EP$), Candidate Reduction ($\CR$), and all possible combinations of all acceleration techniques, on instances of three benchmark problems. 

To ensure a comprehensive comparison, we also conduct all tests using the solvers $\Umod$ and $\Udom$ by~\textcite{sakaue2018}, which achieve the strongest performance reported in the literature, as well as a branch-and-cut ($\BC$) algorithm implemented using Gurobi that solves a binary linear reformulation of \textsc{Submodular Knapsack-Constrained Maximization}. 
Briefly described, the solvers $\Umod$ and $\Udom$ are both \emph{best-first search} algorithms that, given an instance $(I,f,w,B)\in \I$, explore a corresponding basic search tree. Both solvers rely on heuristic functions to determine which node $S$ in a basic search tree should be considered next. $\Umod$ employs a heuristic which is closely related to $\fkh$, while $\Udom$ utilizes a more complex heuristic based on a greedy algorithm for \textsc{Submodular Knapsack-Constrained Maximization}. For details about both algorithms, we refer to~\textcite{sakaue2018}. In the next subsection, we describe the linear reformulation of \textsc{Submodular Knapsack-Constrained Maximization} and the branch-and-cut algorithm used to solve it in detail.

\subsection{A binary linear reformulation of Submodular Knapsack-Constrained Maximization.}

Let $(I,f,w,B)\in\I$. Due to~\textcite{nemhauser1981}, we can reformulate the \textsc{Submodular Knapsack-Constrained Maximization} problem defined by $(I,f,w,B)$ as the following binary linear problem $(\BIP)$:
\begin{align}
\begin{aligned}
    &\max z \\
    &\ \text{s.t. } z \leq f(S)+\sum_{i\in I\setminus S}x_if(i\vert S) \text{ for }S\subseteq I,\\
    &\ \phantom{\text{s.t. }} \sum_{i\in I}x_iw_i \leq B,\\
    &\ \phantom{\text{s.t. }} x_i \in \{0,1\} \text{ for } i\in I.
\end{aligned}
\label{e_prob5}
\end{align}
Since this $\BIP$ has exponentially many constraints (in the number of elements contained in $I$), it is usually impractical to solve it directly.  Instead, we use a branch-and-cut algorithm ($\BC$) based on the following simplified version of Problem~\eqref{e_prob5}:

\begin{align}
    \begin{aligned}
        &\max z \\
    &\ \text{s.t. } z \leq \sum_{i\in I}x_if(\{i\}),\\
    &\ \phantom{\text{s.t. }} \sum_{i\in I}x_iw_i \leq B,\\
    &\ \phantom{\text{s.t. }} x_i \in \{0,1\} \text{ for } i\in I.
    \end{aligned}
    \label{e_prob6}
\end{align}

We solve Problem~\eqref{e_prob5} by first modeling Problem~\eqref{e_prob6} in Gurobi and applying a branch-and-cut approach, dynamically adding the following cut at each integer node $S$ in the search tree to Problem~\eqref{e_prob6}: 

\begin{align*}
    z \leq f(S)+\sum_{i\in I\setminus S}x_if(i\vert S).
\end{align*}
This approach allows us to iteratively refine Problem~\eqref{e_prob6} so that we ultimately obtain an optimal solution to Problem~\eqref{e_prob5}, which in turn corresponds to an optimal solution to $(I,f,w,B)$. 

We implemented all cuts using the \FuncSty{addLazy} function within the Gurobi callback environment at \FuncSty{MIPSOL} locations.

%%%%%%%%%%%%%%%%%%%%%%%%%%%%%%%%%%%%%%%%%%%%%%%%%%%%%%%%%%%%%%%%%%%%%%%%%%%%%%%%%%%%%%%%%%%%%%%%%%%%%%%%%%%%%%%%%%%

\subsection{Experimental setup and data.}

All algorithms are implemented in C++ using the C++17 standard. The code is compiled with GCC version 13.3.0, using the -O2 flag to optimize for efficient machine code. We used Gurobi version 12.0.2 to solve the linear integer problem~\eqref{e_prob2} in each call to $\sn$ in Algorithm~\ref{e_alg1}$^+$, and to implement the branch-and-cut algorithm ($\BC$) that solves the binary linear reformulation of \textsc{Submodular Knapsack-Constrained Maximization}. All experiments were conducted in single-thread mode on a workstation with an AMD EPYC 7552 processor (48 cores, 2.2 GHz base frequency) and $503$ GiB of RAM. The codes for our basic algorithm and the presented acceleration techniques are publicly available at \url{https://github.com/SabineMuench/An-exact-solver-for-submodular-knapsack-problems}.

In line with~\textcite{sakaue2018}, we report computational results for the following well-known benchmark problems:

\begin{enumerate}[label=(\alph*)]
    \item \textsc{Weighted Coverage} ($\cov$). 
    Let $I=\{1,\dots,n\}$ and $E=\{1,\dots,m\}$ be a ground set of elements, with $n,m\in \mathbb{N}$. Further, let $\{E_i\colon E_i\subseteq E, i\in I\}$ be a set of subsets of $E$. Each element $e\in E$ is associated with a value $v_e\geq 0$ and each $i\in I$ costs $w_{i}>0$. The \textsc{Weighted Coverage} problem consists of selecting a subset of $I$ with total costs less than or equal to a given budget $B$ to maximize the weighted coverage function
    \begin{align*}
        f\colon 2^{I}\rightarrow \R_{\geq 0}, X\mapsto \sum_{e\in\bigcup_{i\in X}E_i}v_e.
    \end{align*}
    
    \item \textsc{Facility Location} ($\loc$). Let $I=\{1,\dots,n\}$ be a set of locations and $M=\{1,\dots,m\}$ be a set of customers, with $n,m\in \mathbb{N}$. Each location $i$ is associated with some costs $w_i> 0$, and $v_{ij}\geq 0$ is the benefit that customer $j$ attains from a facility in location $i$. 
    For a subset $X\subseteq I$ of locations, each customer attains the benefit from the facility in the most beneficial location. The objective is to select a subset of locations whose total cost does not exceed a given budget $B\geq 0$, to maximize the total benefit of all customers, which is given by 
    \begin{align*}
        f\colon 2^{I}\rightarrow \R_{\geq 0}, X\mapsto \sum_{j\in M} \max_{i\in X}v_{ij}.
    \end{align*}

    \item \textsc{Bipartite Influence} ($\infe$).
    Let $I=\{1,\dots,n\}$ be a set of sources and $M=\{1,\dots,m\}$ be a set of targets, with $n,m \in \mathbb{N}$. Given a bipartite directed graph $G=(I\cup M, A)$, where $A\subseteq I\times M$ is a set of arcs, we consider an influence maximization problem on $G$. The probability that a target $j\in M$ is activated by a set $X\subseteq I$ is $1-\prod_{i\in X\colon (i,j)\in A} (1-p_i)$, where $0\leq p_i\leq 1$ is the activation probability of source $i\in I$. Every source $i\in I$ is associated with activation costs $w_i> 0$. The goal is to select a set of sources with total costs less than or equal to a given budget $B\geq 0$ to maximize the expected number of activated targets, which is formally defined by
    \begin{align*}
        f\colon 2^{I}\rightarrow \R_{\geq 0}, X\mapsto \sum_{j\in M} \left(1-\prod_{i\in X\colon (i,j)\in A}(1-p_i)\right).
    \end{align*}
\end{enumerate}

\subsubsection{Artificial instances.}
For each of the problems (a)--(c), we construct $100$ artificial instances following the methodology by~\textcite{sakaue2018}. Hence, we set $n=100$ and $m=1000$ for each $n$ and $m$ appearing in the definitions of $\cov$, $\loc$, and $\infe$. In all instances of all problems, we set the budget to $B=1$  and, in each instance of each problem, we draw the costs from the uniform distribution on $\left[0.01,1\right]$. 
Note that~\textcite{sakaue2018} originally drew the costs from the uniform distribution on $[0,1]$. However, to comply with the benchmark problem definitions\ (a)--(c) we adjusted the distribution to avoid the existence of zero-cost elements, which would always be included in any optimal solution.
For $\cov$-instances, we draw the values of the elements in the ground set $E$ from the uniform distribution on $\left[0,1\right]$ and each subset $E_i\subseteq E$ with $i\in \{1,\dots,100\}$, as defined in $\cov$, includes each element of the ground set with probability $0.3$.
For $\loc$-instances, we draw the value $v_{i,j}$, representing the benefit that customer $j$ attains from a facility at location $i$, independently from the uniform distribution over $[0,1]$.
For $\infe$-instances, we randomly draw the activation probabilities for each source from the uniform distribution on $\left[0,1\right]$, and an arc $(i,j)$ from source $i$ to target $j$ is constructed randomly with probability $0.3$.
For all tests, we set the time limit to one hour.

\subsubsection{\texorpdfstring{Discussion of the artificial instances constructed following~\textcite{sakaue2018}.}{}}
We constructed artificial test instances following the methodology used by~\textcite{sakaue2018}. However, we have reservations about these construction methods, as they seem to favor constructing instances that are inherently ``easy'' to solve. Therefore, we further examine how instances of the benchmark problems are constructed by~\textcite{sakaue2018}, outline our concerns with the current approach, and suggest an alternative method for constructing test instances.

We first analyze the construction of $\cov$-instances by~\textcite{sakaue2018} in detail.
In the following, let $I=\{1,\dots,100\}$ and $E=\{1,\dots,1000\}$ be the sets from the definition of $\cov$ as set by~\textcite{sakaue2018}. Since~\textcite{sakaue2018} draw each value of an element in the ground set $E$ from the uniform distribution on $[0,1]$ (in our case from $[0.01,1]$), and each subset $E_i\subseteq E$, with $i\in I$, includes each element from the ground set with probability $0.3$, the expected value of the weighted coverage function, for each $i\in I$, is $150$. Since~\textcite{sakaue2018} draw the costs of each $i\in I$ independently of the weighted coverage function values from the uniform distribution on $[0,1]$, elements $i\in I$ with low costs often have relatively high coverage function values, making them more likely to be included in an optimal solution than elements with higher costs. Hence, an optimal solution for such a $\cov$-instance consists, in expectation, almost exclusively of elements $i\in I$ with low costs. Consequently, elements with greater costs can essentially be ignored when searching for an optimal solution. 
Moreover, this biased structure of the coverage instances favors greedy algorithms for submodular maximization, which are therefore likely to yield objective values that closely approximate the optimal objective value. 
At the same time, this structure causes the fractional knapsack value $\fkh$ to be relatively close to the knapsack value $\kv$ of nearly every node and associated candidate set, considered by the proposed algorithms. This significantly strengthens the used pruning rule. The same holds for the heuristics used in the algorithms $\Umod$ and $\Udom$, to determine which node in a basic search tree should be considered next.
For these reasons, we assume that the $\cov$-instances, which are constructed according to the method by~\textcite{sakaue2018}, are relatively ``easy'' to solve. 

The construction of $\loc$- and $\infe$-instances, as performed by~\textcite{sakaue2018}, exhibits issues similar to the construction of $\cov$-instances. 
In the $\loc$-instances, the benefit of each customer $j$ to be served by a facility in location $i$ is drawn uniformly from $[0,1]$, and all costs are drawn uniformly from $[0,1]$ ($[0.01,1]$ respectively). This introduces a structural bias in the $\loc$-instances, similar to that in the $\cov$-instances, leading to the same effects, since locations with lower costs tend to have disproportionately high objective function values. 
The same applies to the $\infe$-instances, since the activation probabilities of the sources are drawn uniformly from $[0,0.1]$, an arc from source $i$ to target $j$ is constructed with probability $0.3$, and, independently of both the activation probabilities and the arc construction, the activation costs for the sources are drawn from $[0,1]$ ($[0.01,1]$ respectively).

To compensate for these issues in our empirical test, we construct $100$ new instances for each of the three problems as follows:

For $\cov$, we set $n$ and $m$ appearing in the definition to $n=150$ and $m=1000$. We draw the cost of an element $i\in I = \{1,\dots,150\}$ from the uniform distribution on $[0.1,1]$ and set the budget to $B=5$. The value of each element in the ground set $E$ is drawn from the uniform distribution on $[0,1]$ and each subset $E_i\subseteq E$, with $i\in I$, includes each element from the ground set with probability $\frac{w_i}{10}$, where $w_i$ refers to the costs associated with the element $i\in I$.
Hence, the number of elements contained in a subset $E_i$, with $i\in I$, is no longer independent of the costs associated with $i$. If $i\in I$ is associated with higher costs, the subset $E_i\subseteq E$ has a higher probability of containing elements from the ground set $E$ than those subsets $E_j$, with $j\in I$, where $j$ has lower costs. Consequently, in expectation, all elements $i\in I$ exhibit similar relative marginal gains, making the $\cov$-instances more difficult to solve.  

For $\loc$-instances, we set the number of locations to $n=200$ and the number of customers to $m=1000$. We sample the cost of each location $i$ from the uniform distribution on $[0.1,1]$, and for each customer, we draw the benefit they receive from a facility in location $i$ from the uniform distribution on $[0,2w_i]$, where $w_i$ is the cost of location $i$. We set the budget to $B=6$.

For $\infe$-instances, we set the number of targets to $n=150$ and the number of sources to $m=1000$. We draw the costs of each source from the uniform distribution on $[0.1,1]$ and the activation probability from the uniform distribution on $[0,1]$. We construct an arc from source $i$ to target $j$ randomly with probability $\frac{w_i}{5}$, where $w_i$ refers to the costs associated with source $i$. We set the budget to $B=5$.
As before, we set the time limit for all tests to one hour.

\subsubsection{Instances based on real-world data.}
Beyond the tests on artificially constructed instances, we also evaluate all algorithms on $\cov$- and $\infe$-instances generated from real-world data.
For $\cov$, we generated an instance based on real-world data using the Facebook-like Forum Network (weighted by the number of messages) dataset on message exchanges among $899$ users across $522$ topics, as provided by~\textcite{Opsahl2013} and also used by~\textcite{sakaue2018}. This dataset contains information about which users post messages on which topics, as well as a metric based on the number of messages that reflects the importance of a topic to a user.
According to the definition of $\cov$, we define the set of users as $I = \{1, \dots, 899\}$ and the set of topics as $E = \{1, \dots, 522\}$. A topic $e \in E$ is included in the set $E_i$, with $i\in I$, if user $i$ has posted messages on topic $e$ (i.e., user $i$ covers topic $e$). The value of a topic $e$ is defined as the sum of the importance metrics of all users who post on that topic. The cost of a user $i$ is given by the ratio of the number of topics that user $i$ posts messages on multiplied by $100$ and the total number of topics. We set the budget to $B=4$.

To generate an $\infe$-instance based on real-world data, we use the MovieLens 100K dataset by~\textcite{harper2015}, which was also employed by~\textcite{sakaue2018} for this purpose. The dataset contains ratings on a scale from $1$ to $5$, provided by $943$ users for $1682$ movies. Following the definition of $\infe$, we define the set of sources as the set of movies, $I = \{1, \dots, 1682\}$, and the set of targets as the set of users, $M = \{1, \dots, 943\}$. An arc from movie $i$ to user $j$ exists if user $j$ has rated movie $i$. The activation probability $p_i$ of movie $i$ is set to its average rating divided by $10$. The activation cost of a movie $i$ is defined as $2p_i$, and the budget is set to $B =7.5$.
For both real-world data instances, we round costs, benefits, and activation probabilities to
two decimal places.
%%%%%%%%%%%%%%%%%%%%%%%%%%%%%%%%%%%%%%%%%%%%%%%%%%%%%%%%%%%%%%%%%%%%%%%%%%%%%%%%%%%%%%%%%%%%%%%%%%%%%%%%%%%%%%%%%%%

\subsection{\texorpdfstring{Results of the comparison.}{}}
First, we evaluate the performance of the basic solver (Algorithm~\ref{e_alg1}) and its variant with enhanced pruning (Algorithm~\ref{e_alg1}$^+$) on the artificially generated instances. We compare them based on three metrics: the number of instances solved within a 1-hour time limit, the average running time across all solved instances, and the average number of nodes considered during the solution process. Table~\ref{e_tab1} presents the results of these tests on all instances generated according to the methodology described by~\textcite{sakaue2018}, while Table~\ref{e_tab2} shows the results for instances generated using our methodology.
In both tables, the values in parentheses represent the average running time and average number of considered nodes for Algorithm~\ref{e_alg1}, taken over the subset of instances successfully solved by both algorithms. 
If a solver cannot solve any instance of a benchmark problem within the given time limit, we indicate this with '/'.
\newcolumntype{C}[1]{>{\centering\let\newline\\\arraybackslash\hspace{0pt}}m{#1}}

\begin{table}[H]
\centering
\begingroup
\renewcommand{\arraystretch}{0.92}
\begin{tabular}{C{1.cm} C{1.5cm} | C{2.5cm} C{3cm} C{3cm}}
\hline
 &    & solved & time (s) & nodes   \\ \hline
\multirow{2}{*}{$\cov$} & Alg.~\ref{e_alg1}\phantom{$^+$}  & $100$ & $155.11$ $(14.26)$ & $1290204$ $(120071)$ \\ 
& Alg.~\ref{e_alg1}$^+$  & $42$ & $1741.86$ & $95464$\\ \hline
\multirow{2}{*}{$\loc$} & Alg.~\ref{e_alg1}\phantom{$^+$}  & $100$ & $24.05$ $(7.44)$ & $284023$ $(109474)$ \\ 
& Alg.~\ref{e_alg1}$^+$ & $69$ & $1589.86$ & $80491$ \\ \hline
\multirow{2}{*}{$\infe$} & Alg.~\ref{e_alg1}\phantom{$^+$} & $100$ & $14.24$ $(10.63)$ & $59970$ $(43424)$ \\ 
& Alg.~\ref{e_alg1}$^+$  & $96$ & $820.9$ & $36082$ \\ \hline
\end{tabular}
\endgroup
\captionsetup{font=small}
\caption{Number of solved instances, average computation time (s), and average number of processed nodes for Algorithm~\ref{e_alg1} and~\ref{e_alg1}$^+$ on artificial $\cov$-, $\loc$-, and $\infe$-instances generated following~\textcite{sakaue2018}. 
}\label{e_tab1}
\end{table}

\begin{table}[H]
\centering
\begingroup
\renewcommand{\arraystretch}{0.92}
\begin{tabular}{C{1.cm} C{1.5cm} | C{2.5cm} C{3cm} C{3cm}}
\hline
 &    & solved & time (s) & nodes   \\ \hline
\multirow{2}{*}{$\cov$} & Alg.~\ref{e_alg1}\phantom{$^+$}  & $41$ & $1898.53$ & $155778623$  \\ 
& Alg.~\ref{e_alg1}$^+$  & / & / & / \\ \hline
\multirow{2}{*}{$\loc$} & Alg.~\ref{e_alg1}\phantom{$^+$}  & $72$ & $2529.53$ & $7139023$  \\ 
& Alg.~\ref{e_alg1}$^+$  & / & / & / \\ \hline
\multirow{2}{*}{$\infe$} & Alg.~\ref{e_alg1}\phantom{$^+$} & $37$ & $2123.91$ $(157.88)$ & $2408375$ $(141316)$  \\ 
& Alg.~\ref{e_alg1}$^+$  & $2$ & $1977.43$ & $122309$ \\ \hline
\end{tabular}
\endgroup
\captionsetup{font=small}
\caption{Number of solved instances, average computation time (s), and average number of processed nodes for Algorithm~\ref{e_alg1} and~\ref{e_alg1}$^+$ on artificial $\cov$-, $\loc$-, and $\infe$-instances generated following our methodology. 
}\label{e_tab2}
\end{table}  
As mentioned in Section~\ref{e_sec2}, Algorithm~\ref{e_alg1} consistently outperforms Algorithm~\ref{e_alg1}$^+$ in all our tests. It solves more instances, both those generated using the method by~\textcite{sakaue2018} and those generated using our approach, and does so in less time. Although Algorithm~\ref{e_alg1}$^+$ examines fewer nodes, it is overall less effective. The additional pruned nodes in Algorithm~\ref{e_alg1}$^+$ come at the cost of solving an NP-hard problem in each call to $\sn$, which remains computationally expensive even when using state-of-the-art solvers like Gurobi. Our results indicate that this additional effort is not justified: although more nodes are pruned, the associated time cost is too high to improve overall performance compared to Algorithm~\ref{e_alg1}.

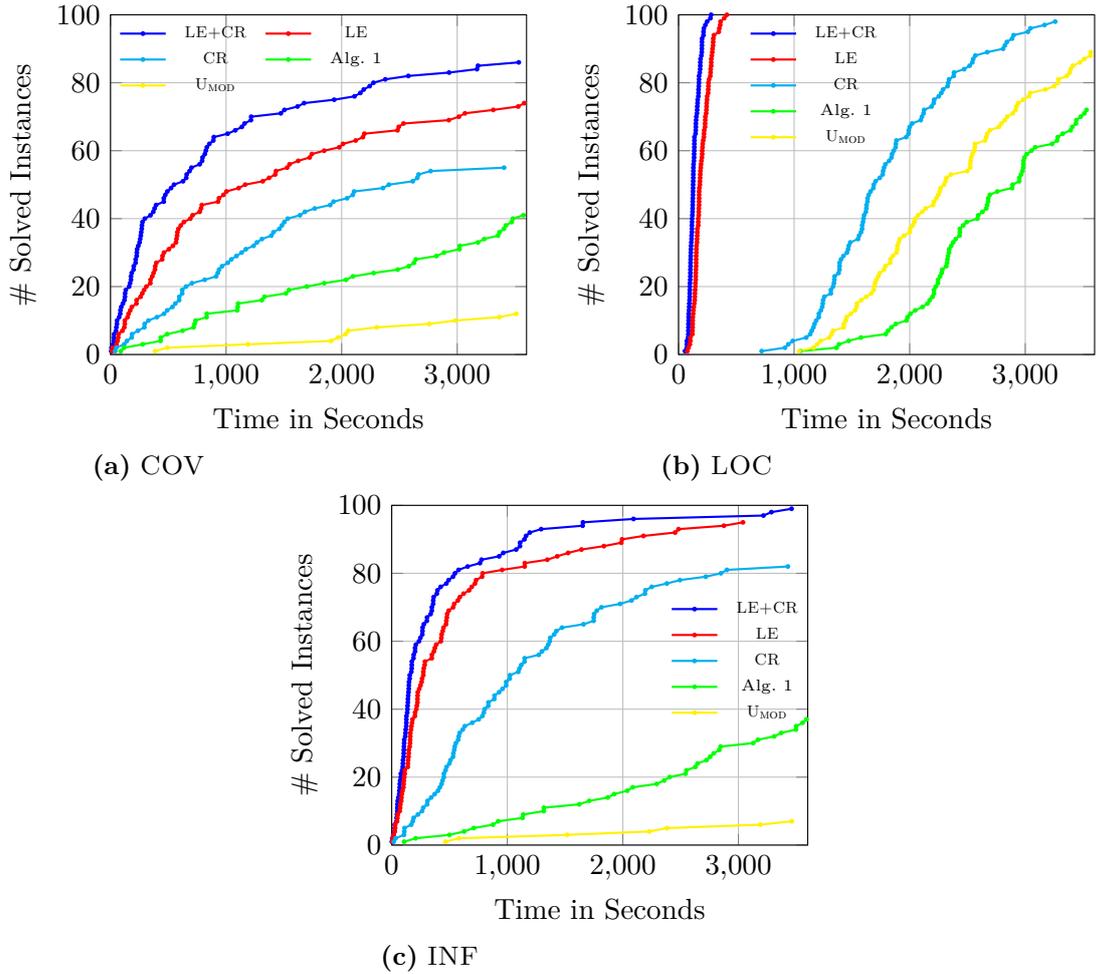
\begin{figure}[H]
\vspace{-1\baselineskip}
\centering
\begin{tikzpicture}
\centering
\pgfplotsset{
    width=0.52\textwidth,
    xlabel={Time in Seconds},
    ylabel={\# Solved Instances},
    grid=major,
}
\begin{axis}[
    name=plot1,
    xmin=0, % X-Achse startet bei 0
    ymin=0, % Y-Achse startet bei 0
    xmax=3600, % Etwas Platz nach rechts
    ymax=100,
    enlargelimits=false,
    legend style={
        at={(1.05,0)}, % Positioniert die Legende außerhalb rechts vom Plot
        anchor=south west,
        legend columns=1, % Vertikale Darstellung der Legende
        column sep=1ex, % Abstand zwischen den Einträgen
        fill=none, % Keine Hintergrundfarbe für die Legende
        draw=none, % Kein Rahmen um die Legende
        font=\small, % Schriftgröße der Legende
    }
]
\addplot[mark=*,
        mark size=0.5pt, % Quadratgröße
        color=Orchid,
        thick
        ] table[x=Sekunden, y=Anzahl] {processed_COV_EP+LE+CRratio150.txt};  
        \addlegendentry{$\LECREP$} 
\addplot[mark=*,
        mark size=0.5pt, % Quadratgröße
        color=orange,
        thick
        ] table[x=Sekunden, y=Anzahl] {processed_COV_EP+LEratio150.txt};  
        \addlegendentry{$\LEEP$} 
\addplot[mark=*,
        mark size=0.5pt, % Quadratgröße
        color=blue,
        thick
        ] table[x=Sekunden, y=Anzahl] {processed_COV_LE+ACRratio150.txt};  
        \addlegendentry{$\LECR$} 
\addplot[mark=*,
        mark size=0.5pt, % Quadratgröße
        color=gray,
        thick
        ] table[x=Sekunden, y=Anzahl] {processed_COV_EP+CRratio150.txt};  
        \addlegendentry{$\EPCR$} 
\addplot[mark=*,
        mark size=0.5pt, % Punktgröße
        color=red,
        thick
        ] table[x=Sekunden, y=Anzahl] {processed_COV_LEratio150.txt};  
        \addlegendentry{$\LE$} 
\addplot[mark=*,
        mark size=0.5pt, % Punktgröße
        color=black,
        thick
        ] table[x=Sekunden, y=Anzahl] {processed_COV_EPratio150.txt};  
        \addlegendentry{$\EP$} 
\addplot[mark=*,
        mark size=0.5pt, % Quadratgröße
        color=cyan,
        thick
        ] table[x=Sekunden, y=Anzahl] {processed_COV_ACRratio150.txt};  
        \addlegendentry{$\CR$}
\addplot[mark=*,
         mark size=0.5pt, % Quadratgröße
         color=Rhodamine,
         thick
        ] table[x=Sekunden, y=Anzahl] {processed_COV_DCO+SUBratio150.txt};  
        \addlegendentry{Alg.~\ref{e_alg1}}
\addplot[mark=*,
        mark size=0.5pt, % Quadratgröße
        color=yellow,
        thick
        ] table[x=Sekunden, y=Anzahl] {processed_COV_Umodratio150.txt};  
        \addlegendentry{$\Umod$}
\end{axis}
 \node at ($(plot1.south west)+(0.5cm,-1.5cm)$) {%
   \parbox{0.42\textwidth}{\centering \small \textbf{(a)} $\cov$\label{e_fig2:COVratio}}};
\end{tikzpicture}
\par
\begin{tikzpicture}
\centering
\pgfplotsset{
    width=0.52\textwidth,
    xlabel={Time in Seconds},
    ylabel={\# Solved Instances},
    grid=major,
}
\begin{axis}[
    name=plot2,
    xmin=0, % X-Achse startet bei 0
    ymin=0, % Y-Achse startet bei 0
    xmax=3600, % Etwas Platz nach rechts
    ymax=100,
    enlargelimits=false,
    legend style={
        at={(1.05,0)}, % Positioniert die Legende außerhalb rechts vom Plot
        anchor=south west,
        legend columns=1, % Vertikale Darstellung der Legende
        column sep=1ex, % Abstand zwischen den Einträgen
        fill=none, % Keine Hintergrundfarbe für die Legende
        draw=none, % Kein Rahmen um die Legende
        font=\small, % Schriftgröße der Legende
    }
]
\addplot[mark=*,
    mark size=0.5pt, % Quadratgröße
    color=Orchid,
    thick
    ] table[x=Sekunden, y=Anzahl] {processed_LOC_EP+LE+CRratio200.txt};  
    \addlegendentry{$\LECREP$} 
\addplot[mark=*,
    mark size=0.5pt, % Quadratgröße
    color=orange,
    thick
    ] table[x=Sekunden, y=Anzahl] {processed_LOC_EP+LEratio200.txt};  
    \addlegendentry{$\LEEP$} 
\addplot[mark=*,
    mark size=0.5pt, % Quadratgröße
    color=blue,
    thick
    ] table[x=Sekunden, y=Anzahl] {processed_LOC_LE+ACRratio200.txt};  
    \addlegendentry{$\LECR$}
\addplot[mark=*,
    mark size=0.5pt, % Punktgröße
    color=red,
    thick
    ] table[x=Sekunden, y=Anzahl] {processed_LOC_LEratio200.txt};  % Beispiel-Datei 1
    \addlegendentry{$\LE$} % Legenden-Eintrag für Linie 1
\addplot[mark=*,
    mark size=0.5pt, % Quadratgröße
    color=green,
    thick] table[x=Sekunden, y=Anzahl] {processed_LOC_BIPratio200.txt};  
    \addlegendentry{$\BC$}  
\addplot[mark=*,
    mark size=0.5pt, % Quadratgröße
    color=gray,
    thick
    ] table[x=Sekunden, y=Anzahl] {processed_LOC_EP+CRratio200.txt};  
    \addlegendentry{$\EPCR$}
\addplot[mark=*,
    mark size=0.5pt, % Quadratgröße
    color=black,
    thick
    ] table[x=Sekunden, y=Anzahl] {processed_LOC_EPratio200.txt};  
    \addlegendentry{$\EP$}
\addplot[mark=*,
    mark size=0.5pt, % Quadratgröße
    color=cyan,
    thick
    ] table[x=Sekunden, y=Anzahl] {processed_LOC_ACRratio200.txt};  
    \addlegendentry{$\CR$}
\addplot[mark=*,
    mark size=0.5pt, % Quadratgröße
    color=yellow,
    thick
    ] table[x=Sekunden, y=Anzahl] {processed_LOC_Umodratio200.txt};  
    \addlegendentry{$\Umod$}
\addplot[mark=*,
    mark size=0.5pt, % Quadratgröße
    color=Rhodamine,
    thick] table[x=Sekunden, y=Anzahl] {processed_LOC_DCO+SUBratio200.txt};  
    \addlegendentry{Alg.~\ref{e_alg1}}   
\end{axis}
\node at ($(plot2.south west)+(0.5cm,-1.5cm)$) {%
   \parbox{0.42\textwidth}{\centering \small \textbf{(b)} $\loc$\label{e_fig2:LOCSakaue}}};
\end{tikzpicture}
\par
\begin{tikzpicture}
\centering
\pgfplotsset{
    width=0.52\textwidth,
    xlabel={Time in Seconds},
    ylabel={\# Solved Instances},
    grid=major,
}
\begin{axis}[
name=plot3, at={($(plot1.south west)+(3.7cm,-2cm)$)}, anchor=north west,
xmin=0, % X-Achse startet bei 0
ymin=0, % Y-Achse startet bei 0
xmax=3600, % Etwas Platz nach rechts
ymax=100,
enlargelimits=false,
legend style={
        at={(1.05,0)}, % Positioniert die Legende außerhalb rechts vom Plot
        anchor=south west,
        legend columns=1, % Vertikale Darstellung der Legende
        column sep=1ex, % Abstand zwischen den Einträgen
        fill=none, % Keine Hintergrundfarbe für die Legende
        draw=none, % Kein Rahmen um die Legende
        font=\small, % Schriftgröße der Legende
    }
]
\addplot[mark=*,
        mark size=0.5pt, % Quadratgröße
        color=Orchid,
        thick
        ] table[x=Sekunden, y=Anzahl] {processed_INF_EP+LE+CRratio150.txt};  
        \addlegendentry{$\LECREP$} 
\addplot[mark=*,
        mark size=0.5pt, % Quadratgröße
        color=blue,
        thick
        ] table[x=Sekunden, y=Anzahl] {processed_INF_LE+ACRratio150.txt};  % Beispiel-Datei 2
        \addlegendentry{$\LECR$}
\addplot[mark=*,
        mark size=0.5pt, % Quadratgröße
        color=orange,
        thick
        ] table[x=Sekunden, y=Anzahl] {processed_INF_EP+LEratio150.txt};  
        \addlegendentry{$\LEEP$} 
\addplot[mark=*,
        mark size=0.5pt, % Punktgröße
        color=red,
        thick
        ] table[x=Sekunden, y=Anzahl] {processed_INF_LEratio150.txt};  % Beispiel-Datei 1
        \addlegendentry{$\LE$} % Legenden-Eintrag für Linie 1  
\addplot[mark=*,
        mark size=0.5pt, % Quadratgröße
        color=gray,
        thick
        ] table[x=Sekunden, y=Anzahl] {processed_INF_EP+CRratio150.txt};  
        \addlegendentry{$\EPCR$}
\addplot[
        mark=*,
        mark size=0.5pt, % Quadratgröße
        color=cyan,
        thick
        ] table[x=Sekunden, y=Anzahl] {processed_INF_ACRratio150.txt};  
        \addlegendentry{$\CR$}
\addplot[mark=*,
        mark size=0.5pt, % Quadratgröße
        color=black,
        thick
        ] table[x=Sekunden, y=Anzahl] {processed_INF_EPratio150.txt};  
        \addlegendentry{$\EP$}
\addplot[
        mark=*,
        mark size=0.5pt, % Quadratgröße
        color=Rhodamine,
        thick
        ] table[x=Sekunden, y=Anzahl] {processed_INF_DCO+SUBratio150.txt};  
        \addlegendentry{Alg.~\ref{e_alg1}}
\addplot[mark=*,
        mark size=0.5pt, % Quadratgröße
        color=yellow,
        thick
        ] table[x=Sekunden, y=Anzahl] {processed_INF_Umodratio150.txt};  
        \addlegendentry{$\Umod$}
\addplot[mark=*,
        mark size=0.5pt, % Quadratgröße
        color=ForestGreen,
        thick
        ] table[x=Sekunden, y=Anzahl] {processed_INF_LEindexratio150.txt};  
        \addlegendentry{$\LEg$}
\end{axis}
\node at ($(plot3.south west)+(0.5cm,-1.5cm)$) {%
   \parbox{0.42\textwidth}{\centering \small \textbf{(c)} $\infe$\label{e_fig2:INFSakaue}}};
\end{tikzpicture}
\captionsetup{font=small}
\caption{Number of solved instances for all solvers on artificial $\cov$-, $\loc$-, and $\infe$- instances generated following our method.\label{e_fig2}}
\end{figure}

Next, we evaluate the performance of the accelerated versions of Algorithm~\ref{e_alg1} on the artificially generated instances. Specifically, we compare Lazy Evaluations, Early Pruning, Candidate Reduction, and all possible combinations of the acceleration techniques against Algorithm~\ref{e_alg1}, the solvers $\Umod$ and $\Udom$ by~\textcite{sakaue2018}, and the branch-and-cut algorithm $\BC$. As before, the comparison is based on the number of instances solved within a 1-hour time limit, the average running time, and the average number of nodes considered. The results of our tests are summarized in Tables~\ref{e_tab3} and~\ref{e_tab4}, and Figures~\ref{e_fig1} and~\ref{e_fig2}.
Notice that we deferred the results for the combinations $\LEgCR$, $\LEgEP$, and $\LEgCREP$ to the Appendix, as $\LEg$ performed poorly compared to both Algorithm~\ref{e_alg1} and the other acceleration techniques, and its combinations with $\EP$ and $\CR$ did not change this substantially. Moreover, these results provide only limited further insight beyond the comparison with the combinations of $\LE$ with $\EP$, and $\CR$.

Both $\Umod$ and $\Udom$ are decisively outperformed by all acceleration techniques we propose except $\LEg$. Despite considering significantly fewer nodes on average, $\Udom$ solves far fewer instances of all problem types within the time limit and has a higher running time for the instances it solves. In particular, $\Udom$ fails to solve any of the instances generated by our method within the given time limit. $\Umod$ performs slightly better than $\Udom$, but it is significantly slower than Algorithm~\ref{e_alg1} across all problem instances, except for the $\loc$ instances. However, overall Algorithm~\ref{e_alg1} and the proposed acceleration techniques, except $\LEg$, are the stronger solvers, as they solve more instances within the given time limit across all problem types and achieve lower average running times.

The branch-and-cut algorithm $(\BC)$ does not solve any $\cov$- or $\infe$-instance constructed following our method within the given time limit. In contrast, it solves all $\loc$-instances constructed according to our method and considers far fewer nodes than Algorithm~\ref{e_alg1}, the presented acceleration techniques, and their combinations. 
Overall, $\BC$ is outperformed by combinations of the acceleration techniques $\LE$, $\CR$, and $\EP$, and in some cases also by their individual application. On average, $\LECREP$ solves all instances of all problems faster than $\BC$. In addition, the $\loc$ and $\infe$ instances constructed according to the method by~\textcite{sakaue2018} are solved significantly faster by all combinations of $\LE$, $\CR$, and $\EP$, as well as by the acceleration techniques applied individually, than by $\BC$. For example, $\LECR$ solves the $\loc$ instances generated following~\textcite{sakaue2018} $42$ times faster than $\BC$.

Incorporating the presented acceleration techniques into Algorithm~\ref{e_alg1} generally results in a substantial improvement in the performance of Algorithm~\ref{e_alg1}. The only exception is the incorporation of Lazy Evaluations with the greedy decision rule ($\LEg$), which has a negative influence on the average running time of Algorithm~\ref{e_alg1}. In contrast, incorporating Lazy Evaluations with the average decision rule ($\LE$) yields the greatest running time reduction among all acceleration techniques.
\begin{table}[H]
\centering
\begingroup
\renewcommand{\arraystretch}{0.92}
\begin{tabular}{C{1.cm} C{2.7cm} | C{2.5cm} C{2.5cm} C{3cm}}
\hline
&    & solved & time (s) & nodes  \\ \hline
\multirow{12}{*}{$\cov$} & Alg.~\ref{e_alg1} & $100$ & $155.11$ & $1290204$  \\
& $\LE$     & $100$ & $49.77$ & $1582353$  \\
& $\LEg$    & $99$  & $186.81$ & $8325021$   \\
& $\EP$     & $100$ & $103.15$  & $1290204$  \\
& $\CR$     & $100$ & $139.97$ & $416637$ \\  
& $\LECR$   & $100$ & $59.86$ & $620550$  \\
& $\LEEP$   & $100$ & $50.12$ & $1582353$ \\
& $\EPCR$   & $100$ & $59.23$& $416637$   \\
& $\LECREP$ & $100$ & $33.64$ & $620570$ \\
& $\Umod$   & $98$ & $264.68$ & $29166$ \\
& $\Udom$   & $96$ & $409.29$ & $70221$ \\
& $\BC$ & $100$ & $39.39$ & $232389$ \\
\hline
\multirow{12}{*}{$\loc$} & Alg.~\ref{e_alg1} & $100$ & $24.05$ & $284023$   \\
& $\LE$     & $100$ & $5.48$ & $399478$  \\
& $\LEg$    & $100$ & $86.21$ & $6780993$\\
& $\EP$     & $100$ & $12.3$ & $284023$ \\
& $\CR$     & $100$ & $7.68$ & $84952$   \\  
& $\LECR$   & $100$ & $2.29$ & $141541$  \\
& $\LEEP$   & $100$ & $5.07$ & $399478$  \\
& $\EPCR$   & $100$ & $6.06$ & $84952$  \\
& $\LECREP$ & $100$ & $3.14$   & $141541$ \\
& $\Umod$   & $100$ & $21.95$ & $7849$   \\
& $\Udom$   & $100$ & $49.83$ & $15790$  \\
& $\BC$ & $100$ & $97.04$ & $47002$ \\
\hline
\multirow{12}{*}{$\infe$} & Alg.~\ref{e_alg1} & $100$ & $14.24$ & $59970$  \\
& $\LE$     & $100$ & $2.01$ & $83075$   \\
& $\LEg$    & $100$ & $39.68$& $1270941$ \\
& $\EP$     & $100$ & $5.06$ & $59970$   \\
& $\CR$     & $100$ & $3.54$ & $16972$   \\  
& $\LECR$   & $100$ & $1.05$ & $26933$   \\
& $\LEEP$   & $100$ & $2.14$ & $83075$   \\
& $\EPCR$   & $100$ & $2.09$ & $16972$   \\
& $\LECREP$ & $100$ & $0.91$ & $26933$   \\
& $\Umod$   & $100$ & $23.42$ & $2202$   \\
& $\Udom$   & $100$ & $166.03$ & $19112$ \\
& $\BC$ & $100$ & $11.07$ & $7895$ \\
\hline
\end{tabular}
\endgroup
\captionsetup{font=small}
\caption{Number of solved instances, average computation time (s), and average number of processed nodes for all solvers on artificial $\cov$-, $\loc$-, and $\infe$-instances generated following the method by~\textcite{sakaue2018}.}\label{e_tab3}
\end{table}

\begin{table}[H]
\centering
\begingroup
\renewcommand{\arraystretch}{0.92}
\begin{tabular}{C{1.cm} C{2.7cm} | C{2.5cm} C{2.5cm} C{3cm}}
\hline
&    & solved & time (s) & nodes  \\ \hline
\multirow{12}{*}{$\cov$} & Alg.~\ref{e_alg1} & $41$ & $1898.53$ & $15577862$ \\
& $\LE$     & $74$ & $1013.24$ & $98520775$ \\
& $\LEg$    & / & / & / \\
& $\EP$     & $61$ & $1111.51$ & $27750444$  \\
& $\CR$     & $55$ & $1177.77$ & $8761085$ \\  
& $\LECR$   & $86$ & $756.87$  & $57276858$ \\
& $\LEEP$   & $89$ & $777.81$  & $164716670$ \\
& $\EPCR$   & $83$ & $836.71$ & $21123028$ \\
& $\LECREP$ & $93$ & $664.16$  & $79446441$  \\
& $\Umod$   & $12$ & $2119.82$ & $96082$ \\
& $\Udom$   & / & / & /  \\
& $\BC$    & / & / & /  \\
\hline
\multirow{12}{*}{$\loc$} & Alg.~\ref{e_alg1} & $72$ & $2529.53$ & $7139023$ \\
& $\LE$     &$100$ & $202.27$ & $8828172$ \\
& $\LEg$    & / & / & / \\
& $\EP$     &$100$ & $1093.78$& $8239881$ \\
& $\CR$     &$98$  & $1818.49$& $5713070$ \\  
& $\LECR$   &$100$ & $137.18$ & $6672304$ \\
& $\LEEP$   &$100$ & $135.6$  & $8828228$ \\
& $\EPCR$   &$100$ & $764.78$ & $5774286$ \\
& $\LECREP$ &$100$ & $133.26$ & $6672347$ \\
& $\Umod$   & $90$ & $2293.81$ &$63613$   \\
& $\Udom$   & / & / & / \\
& $\BC$    & $100$ & $217.49$ & $49131$ \\
\hline
\multirow{12}{*}{$\infe$} & Alg.~\ref{e_alg1} & $37$ & $2123.91$ & $2408375$   \\
& $\LE$     & $95$ & $522.54$ & $13698321$  \\
& $\LEg$    & $1$  & $2648.4$ & $109731524$ \\
& $\EP$     & $81$ & $1196.06$ & $5779142$  \\
& $\CR$     & $82$ & $1029.85$ & $2105902$  \\  
& $\LECR$   & $99$ & $353.73$  & $7993367$  \\
& $\LEEP$   & $96$ & $388.63$  & $146076978$ \\
& $\EPCR$   & $93$ & $747.07$ & $3072641$ \\
& $\LECREP$ & $100$ & $320.16$ & $8990674$  \\
& $\Umod$   & $9$  & $2205.62$ & $25519$    \\
& $\Udom$   & / & / & /  \\
& $\BC$    & / & / & /  \\
\hline
\end{tabular}
\endgroup
\captionsetup{font=small}
\caption{Number of solved instances, average computation time (s), and average number of processed nodes for all solvers on artificial $\cov$-, $\loc$-, and $\infe$-instances generated following our method.}\label{e_tab4}
\end{table}
Although the running time improvements achieved by $\EP$ and $\CR$ are not as large as those achieved by $\LE$, they still represent a consistent speedup of Algorithm~\ref{e_alg1}.

The combination of acceleration techniques results in an additional reduction in running time beyond that achieved by the individually applied techniques for instances of all three problems generated using our method. Therefore, $\LECREP$ is the strongest solver for all problems on these instances. Specifically, $\LECREP$ solves $52\%$ more of the $\cov$-instances, $28\%$ more of the $\loc$-instances, and $63\%$ more of the $\infe$-instances than Algorithm~\ref{e_alg1}.
Considering the instances constructed according to the method by~\textcite{sakaue2018}, the combination of acceleration techniques does not always yield an additional reduction in running time beyond that achieved by the individually applied techniques. Specifically, the $\cov$-instances are solved more slowly by $\LECR$ and $\LEEP$ than by $\LE$, the $\loc$-instances more slowly by $\LECREP$ than by $\LECR$, and the $\infe$-instances more slowly by $\LEEP$ than by $\LE$. Hence, combining more acceleration techniques does not necessarily result in greater acceleration. 

However, $\LECREP$ remains the strongest solver on the $\cov$- and $\infe$-instances constructed according to the method by~\textcite{sakaue2018}, solving the $\cov$-instances four times faster than Algorithm~\ref{e_alg1} and the $\infe$-instances even $15$ times faster. The $\loc$-instances generated following the method by~\textcite{sakaue2018} are solved fastest by $\LECR$; specifically, they are solved $10$ times faster by $\LECR$ than by Algorithm~\ref{e_alg1}.

Interestingly, $\LECR$ considers more nodes on average than $\CR$, but fewer than $\LE$ across all instances of all problems. There are two reasons for this behavior. First, $\LECR$, unlike $\CR$, uses a pruning rule based on the lazy fractional knapsack value rather than the fractional knapsack value. As a direct consequence, $\LECR$ prunes at most as many nodes as $\CR$. Second, when constructing the reduction set $R(S)$ of a node $S \neq \emptyset$ in $\LECR$, we do not always use the exact marginal gain $f(c \vert S)$ of an element $c \in C(S)$. Instead, in accordance with Lazy Evaluations, we use the upper bound $u_{S}(c)$, where $u_{S}$ is the lazy evaluation scheme in $S$.
Since $u_{S}(c)+\lfkh(S,C(S)\setminus\{c\})\geq f(c\vert S)+\fkh(S,C(S)\setminus\{c\})$, we have

\begin{align*}
    R(S) &= \{c\in C(S)\colon u_{S}(c)+\lfkh(S,C(S)\setminus\{c\})\leq s^*\}\\
    &\subseteq  \{c\in C(S)\colon f(c\vert S)+\fkh(S,C(S)\setminus\{c\})\leq s^*\}
\end{align*}
for the reduction set $R(S)$ in each call $\sn(S, C(S), t^*)$ with $S\neq \emptyset$ in $\LECR$.
Hence, it follows directly that $\LECR$ excludes fewer nodes from consideration than $\CR$ and consequently explores more nodes. Since $\LE$ uses the same pruning rule as $\LECR$ but does not exclude the nodes in $R(S)$ from consideration, it naturally considers more nodes than $\LECR.$

Additionally, the tests empirically prove that the instances constructed according to the method by~\textcite{sakaue2018} are generally ``easy'' to solve, since nearly all solvers we compare were able to solve all of these instances, and the average running times, particularly for the acceleration techniques, were close to each other. Both $\Umod$ and $\Udom$, as well as the algorithms proposed by us and $\BC$, tend to solve fewer instances or require more time when applied to instances generated using our method compared to those generated following~\textcite{sakaue2018}.
\begin{figure}[H]
\vspace{-1\baselineskip}
\centering
\begin{tikzpicture}
\centering
\pgfplotsset{
    width=0.52\textwidth,
    xlabel={Time in Seconds},
    ylabel={\# Solved Instances},
    grid=major,
}
\begin{axis}[
    name=plot1,
    xmin=0, % X-Achse startet bei 0
    ymin=0, % Y-Achse startet bei 0
    xmax=3203, % Etwas Platz nach rechts
    ymax=100,
    enlargelimits=false,
    legend style={
        at={(1.05,0.37)}, % Positioniert die Legende außerhalb rechts vom Plot
        anchor=south west,
        legend columns=2, % Vertikale Darstellung der Legende
        column sep=1ex, % Abstand zwischen den Einträgen
        fill=none, % Keine Hintergrundfarbe für die Legende
        draw=none, % Kein Rahmen um die Legende
        font=\small, % Schriftgröße der Legende
    }
]
\addplot[mark=*,
         mark size=0.5pt, % Quadratgröße
         color=green,
         thick
        ] table[x=Sekunden, y=Anzahl] {processed_COV_BIPSakaue.txt};  
        \addlegendentry{$\BC$}
\addplot[mark=*,
         mark size=0.5pt, % Quadratgröße
         color=Orchid,
         thick
        ] table[x=Sekunden, y=Anzahl] {processed_COV_EP+LE+CRSakaue.txt};  
        \addlegendentry{$\LECREP$}
\addplot[mark=*,
        mark size=0.5pt, % Punktgröße
        color=red,
        thick
        ] table[x=Sekunden, y=Anzahl] {processed_COV_LESakaue.txt};  % Beispiel-Datei 1
        \addlegendentry{$\LE$} % Legenden-Eintrag für Linie 1
\addplot[mark=*,
         mark size=0.5pt, % Quadratgröße
         color=orange,
         thick
        ] table[x=Sekunden, y=Anzahl] {processed_COV_EP+LESakaue.txt};  
        \addlegendentry{$\LEEP$}
\addplot[mark=*,
        mark size=0.5pt, % Quadratgröße
        color=blue,
        thick
        ] table[x=Sekunden, y=Anzahl] {processed_COV_LE+ACRSakaue.txt};  % Beispiel-Datei 2
        \addlegendentry{$\LECR$} % Legenden-Eintrag für Linie 2 
\addplot[mark=*,
         mark size=0.5pt, % Quadratgröße
         color=gray,
         thick
        ] table[x=Sekunden, y=Anzahl] {processed_COV_EP+CRSakaue.txt};  
        \addlegendentry{$\EPCR$}
\addplot[mark=*,
        mark size=0.5pt, % Quadratgröße
        color=black,
        thick
        ] table[x=Sekunden, y=Anzahl] {processed_COV_EPSakaue.txt};  
        \addlegendentry{$\EP$}
\addplot[mark=*,
        mark size=0.5pt, % Quadratgröße
        color=cyan,
        thick
        ] table[x=Sekunden, y=Anzahl] {processed_COV_ACRSakaue.txt};  
        \addlegendentry{$\CR$}
\addplot[mark=*,
         mark size=0.5pt, % Quadratgröße
         color=Rhodamine,
         thick
        ] table[x=Sekunden, y=Anzahl] {processed_COV_DCO+SUBSakaue.txt};  
        \addlegendentry{Alg.~\ref{e_alg1}}
\addplot[mark=*,
        mark size=0.5pt, % Quadratgröße
        color=ForestGreen,
        thick
        ] table[x=Sekunden, y=Anzahl] {processed_COV_LEindexSakaue.txt};  
        \addlegendentry{$\LEg$}
\addplot[mark=*,
        mark size=0.5pt, % Quadratgröße
        color=yellow,
        thick
        ] table[x=Sekunden, y=Anzahl] {processed_COV_UmodSakaue.txt};  
        \addlegendentry{$\Umod$}
\addplot[mark=*,
         mark size=0.5pt, % Quadratgröße
         color=brown,
         thick
        ] table[x=Sekunden, y=Anzahl] {processed_COV_UdomSakaue.txt};  
        \addlegendentry{$\Udom$}
\end{axis}
\node at ($(plot1.south west)+(0.5cm,-1.5cm)$) {%
   \parbox{0.42\textwidth}{\centering \small \textbf{(a)} $\cov$\label{e_fig1:COVSakaue}}};
\end{tikzpicture}
\par
%\vspace{0.5cm}
\begin{tikzpicture}
\centering
\pgfplotsset{
    width=0.52\textwidth,
    xlabel={Time in Seconds},
    ylabel={\# Solved Instances},
    grid=major,
}
\begin{axis}[
    name=plot2, % at={($(plot1.north east)+(2cm,0)$)}, anchor=north west,
    xmin=0, % X-Achse startet bei 0
    ymin=0, % Y-Achse startet bei 0
    xmax=400, % Etwas Platz nach rechts
    ymax=100,
    enlargelimits=false,
    legend style={
        at={(1.05,0.37)}, % Positioniert die Legende außerhalb rechts vom Plot
        anchor=south west,
        legend columns=2, % Vertikale Darstellung der Legende
        column sep=1ex, % Abstand zwischen den Einträgen
        fill=none, % Keine Hintergrundfarbe für die Legende
        draw=none, % Kein Rahmen um die Legende
        font=\small, % Schriftgröße der Legende
    }
]
\addplot[mark=*,
        mark size=0.5pt, % Quadratgröße
        color=blue,
        thick
        ] table[x=Sekunden, y=Anzahl] {processed_LOC_LE+ACRSakaue.txt};  
        \addlegendentry{$\LECR$}
\addplot[mark=*,
        mark size=0.5pt, % Quadratgröße
        color=Orchid,
        thick
        ] table[x=Sekunden, y=Anzahl] {processed_LOC_EP+LE+CRSakaue.txt};  
        \addlegendentry{$\LECREP$}
\addplot[mark=*,
        mark size=0.5pt, % Punktgröße
        color=red,
        thick
        ] table[x=Sekunden, y=Anzahl] {processed_LOC_LESakaue.txt};  % Beispiel-Datei 1
        \addlegendentry{$\LE$} % Legenden-Eintrag für Linie 1
\addplot[mark=*,
        mark size=0.5pt, % Quadratgröße
        color=orange,
        thick
        ] table[x=Sekunden, y=Anzahl] {processed_LOC_EP+LESakaue.txt};  
        \addlegendentry{$\LEEP$}
\addplot[mark=*,
        mark size=0.5pt, % Quadratgröße
        color=gray,
        thick
        ] table[x=Sekunden, y=Anzahl] {processed_LOC_EP+CRSakaue.txt};  
        \addlegendentry{$\EPCR$}
\addplot[mark=*,
        mark size=0.5pt, % Quadratgröße
        color=cyan,
        thick
        ] table[x=Sekunden, y=Anzahl] {processed_LOC_ACRSakaue.txt};  
        \addlegendentry{$\CR$}
\addplot[mark=*,
        mark size=0.5pt, % Quadratgröße
        color=black,
        thick
        ] table[x=Sekunden, y=Anzahl] {processed_LOC_EPSakaue.txt};  
        \addlegendentry{$\EP$}
\addplot[mark=*,
        mark size=0.5pt, % Quadratgröße
        color=yellow,
        thick
        ] table[x=Sekunden, y=Anzahl] {processed_LOC_UmodSakaue.txt};  
        \addlegendentry{$\Umod$}
\addplot[mark=*,
        mark size=0.5pt, % Quadratgröße
        color=Rhodamine,
        thick
        ] table[x=Sekunden, y=Anzahl] {processed_LOC_DCO+SUBSakaue.txt};  
        \addlegendentry{Alg.~\ref{e_alg1}}   
\addplot[
        mark=*,
        mark size=0.5pt, % Quadratgröße
        color=brown,
        thick
        ] table[x=Sekunden, y=Anzahl] {processed_LOC_UdomSakaue.txt};  
        \addlegendentry{$\Udom$}
\addplot[mark=*,
        mark size=0.5pt, % Quadratgröße
        color=green,
        thick
        ] table[x=Sekunden, y=Anzahl] {processed_LOC_BIPSakaue.txt};  
        \addlegendentry{$\BC$}
\addplot[mark=*,
        mark size=0.5pt, % Quadratgröße
        color=ForestGreen,
        thick
        ] table[x=Sekunden, y=Anzahl] {processed_LOC_LEindexSakaue.txt};  
        \addlegendentry{$\LEg$}
\end{axis}
\node at ($(plot2.south west)+(0.5cm,-1.5cm)$) {%
  \parbox{0.42\textwidth}{\centering \small \textbf{(b)} $\loc$\label{e_fig1:LOCSakaue}}};
\end{tikzpicture}
\par
%\vspace{0.5cm}
\begin{tikzpicture}
\centering
\pgfplotsset{
    width=0.52\textwidth,
    xlabel={Time in Seconds},
    ylabel={\# Solved Instances},
    grid=major,
}
\begin{axis}[
    name=plot3,% at={($(plot1.south west)+(3.7cm,-2cm)$)}, anchor=north west,
    xmin=0, % X-Achse startet bei 0
    ymin=0, % Y-Achse startet bei 0
    xmax=2302, % Etwas Platz nach rechts
    ymax=100,
    enlargelimits=false,
    legend style={
        at={(1.05,0.37)}, % Positioniert die Legende außerhalb rechts vom Plot
        anchor=south west,
        legend columns=2, % Vertikale Darstellung der Legende
        column sep=1ex, % Abstand zwischen den Einträgen
        fill=none, % Keine Hintergrundfarbe für die Legende
        draw=none, % Kein Rahmen um die Legende
        font=\small, % Schriftgröße der Legende
    }
]
\addplot[mark=*,
        mark size=0.5pt, % Quadratgröße
        color=Orchid,
        thick
        ] table[x=Sekunden, y=Anzahl] {processed_INF_EP+LE+CRSakaue.txt};  
        \addlegendentry{$\LECREP$}
\addplot[mark=*,
        mark size=0.5pt, % Quadratgröße
        color=blue,
        thick
        ] table[x=Sekunden, y=Anzahl] {processed_INF_LE+ACRSakaue.txt};  % Beispiel-Datei 2
        \addlegendentry{$\LECR$}
\addplot[mark=*,
        mark size=0.5pt, % Quadratgröße
        color=orange,
        thick
        ] table[x=Sekunden, y=Anzahl] {processed_INF_EP+LESakaue.txt};  
        \addlegendentry{$\LEEP$}
\addplot[mark=*,
        mark size=0.5pt, % Punktgröße
        color=red,
        thick
        ] table[x=Sekunden, y=Anzahl] {processed_INF_LESakaue.txt};  % Beispiel-Datei 1
        \addlegendentry{$\LE$} % Legenden-Eintrag für Linie 1
\addplot[mark=*,
        mark size=0.5pt, % Quadratgröße
        color=gray,
        thick
        ] table[x=Sekunden, y=Anzahl] {processed_INF_EP+CRSakaue.txt};  
        \addlegendentry{$\EPCR$}  
\addplot[mark=*,
        mark size=0.5pt, % Quadratgröße
        color=cyan,
        thick
        ] table[x=Sekunden, y=Anzahl] {processed_INF_ACRSakaue.txt};  
        \addlegendentry{$\CR$}
\addplot[mark=*,
        mark size=0.5pt, % Quadratgröße
        color=black,
        thick
        ] table[x=Sekunden, y=Anzahl] {processed_INF_EPSakaue.txt};  
        \addlegendentry{$\EP$}
\addplot[mark=*,
        mark size=0.5pt, % Quadratgröße
        color=green,
        thick
        ] table[x=Sekunden, y=Anzahl] {processed_INF_BIPSakaue.txt};  
        \addlegendentry{$\BC$}
\addplot[mark=*,
        mark size=0.5pt, % Quadratgröße
        color=Rhodamine,
        thick
        ] table[x=Sekunden, y=Anzahl] {processed_INF_DCO+SUBSakaue.txt};  
        \addlegendentry{Alg.~\ref{e_alg1}}
\addplot[mark=*,
        mark size=0.5pt, % Quadratgröße
        color=yellow,
        thick
        ] table[x=Sekunden, y=Anzahl] {processed_INF_UmodSakaue.txt};  
        \addlegendentry{$\Umod$}
\addplot[mark=*,
        mark size=0.5pt, % Quadratgröße
        color=ForestGreen,
        thick
        ] table[x=Sekunden, y=Anzahl] {processed_INF_LEindexSakaue.txt};  
        \addlegendentry{$\LEg$}
\addplot[mark=*,
        mark size=0.5pt, % Quadratgröße
        color=brown,
        thick
        ] table[x=Sekunden, y=Anzahl] {processed_INF_UdomSakaue.txt};  
        \addlegendentry{$\Udom$}
\end{axis}
\node at ($(plot3.south west)+(0.5cm,-1.5cm)$) {%
   \parbox{0.42\textwidth}{\centering \small \textbf{(c)} $\infe$\label{e_fig1:INFSakaue}}};
\end{tikzpicture}
\captionsetup{font=small}
\caption{Number of solved instances for all solvers on artificial $\cov$-, $\loc$-, and $\infe$- instances generated following the method by~\textcite{sakaue2018}.\label{e_fig1}}
\end{figure}

Finally, we compare all solvers on the two instances derived from real-world data. As before, we evaluate both the running time (in seconds) and the number of nodes explored. For these experiments, we increase the time limit to $24$ hours. Table~\ref{e_tab5} presents the results of these experiments. 

\begin{table}[H]
\centering
\begin{tabular}{C{1.cm} C{2.7cm} | C{2.5cm} C{3cm}}
\hline
&   & time (s) & nodes  \\ \hline
\multirow{12}{*}{$\cov$} & Alg.~\ref{e_alg1} & $8136.91$ & $123795086$ \\
& $\LE$     &  $12122.4$& $345855983$  \\
& $\LEg$    &  $69649.5$& $1073706857$ \\
& $\EP$     &  $317.63$ & $123795086$  \\
& $\CR$     &  $199.29$ & $9691498$    \\  
& $\LECR$   &  $203.34$ & $18806204$   \\
& $\LEEP$   &  $626.08$ & $345855983$  \\
& $\EPCR$   &  $49.3$   & $9691498$    \\
& $\LECREP$ &  $172.58$ & $18807255$   \\
& $\Umod$   &  $10360.2$& $182209$     \\
& $\Udom$   & / & /  \\
& $\BC$     & $3.1$ & $4369$ \\
\hline
\multirow{12}{*}{$\infe$} & Alg.~\ref{e_alg1} & / & / \\
& $\LE$     & $2546.69$ & $22648276$  \\
& $\LEg$    & / & / \\
& $\EP$     & $2375.81$ & $13576915$  \\
& $\CR$     & $18319.5$ & $4005760$   \\  
& $\LECR$   & $502.84$  & $5498329$   \\
& $\LEEP$   & $375.05$  & $22648276$  \\
& $\EPCR$   & $841.34$  & $4005348$   \\
& $\LECREP$ & $182.34$  & $5498111$   \\
& $\Umod$   & $82694.2$& $8167$ \\
& $\Udom$   & / & / \\
& $\BC$     & / & / \\
\hline
\end{tabular}
\captionsetup{font=small}
\caption{Computation time (in seconds) and number of processed nodes for instances based on real-world data.}\label{e_tab5}
\end{table}
$\BC$ is by far the fastest solver for the $\cov$-instance based on real-world data, whereas the $\infe$-instance cannot be solved by $\BC$ within the given time limit. The superiority of $\BC$ on the $\cov$-instance based on real-world data is due to its special structure. In this instance, some topics have very high values and are covered by users who post messages on only a few other topics, leading to low costs associated with these users. These users are included in every optimal solution to this $\cov$-instance, and in addition, $\BC$ already sets the corresponding decision variables $x_i$ equal to $1$ at an early stage of the search tree.
Consequently, $\BC$ adds strong cuts to Problem~\eqref{e_prob6} early on, allowing it to find an optimal solution quickly.

Interestingly, both $\LE$ and $\LEg$ perform poorly on the $\cov$-instance based on real-world data, while $\CR$ and $\EP$ provide significant accelerations over Algorithm~\ref{e_alg1}. Consistent with these results, $\EPCR$ solves the $\cov$-instances fastest after $\BC$. 
Notably, $\Umod$ can solve the $\infe$-instance within the given time limit of $24$ hours, unlike Algorithm~\ref{e_alg1}. However, except for $\LEg$, all of our acceleration techniques are very successful on the $\infe$-instance based on real-world data. $\LECREP$ is the fastest solver for this instance. In general, our methods -- particularly the combinations of the acceleration techniques presented -- solve most of the instances considered, on average, substantially faster than $\Umod$ and $\Udom$ by~\textcite{sakaue2018}, and the branch-and-cut algorithm $\BC$.
%%%%%%%%%%%%%%%%%%%%%%%%%%%%%%%%%%%%%%%%%%%%%%%%%%%%%%%%%%%%%%%%%%%%%%%%%%%%%%%%%%%%%%%%%%%%%%%%%%%%%%%%%%%%%%%%%

\subsubsection{\texorpdfstring{Brief discussion of experimental results for $\Umod$ and $\Udom$.}{}}
\textcite{sakaue2018} state that $\Udom$ substantially outperforms $\Umod$ in all aspects, which is supported by the experimental results in their paper.
However, we were unable to reproduce the positive test results reported for $\Udom$. In our tests, $\Umod$ proved to be the stronger of the two solvers. 

Notice that we conducted the test of $\Umod$ and $\Udom$ using reimplementations of both algorithms, as the original implementations were neither publicly available nor provided upon request. A detailed discussion of our implementations and the description of $\Udom$ by~\textcite{sakaue2018} is deferred to the Appendix. 

Interestingly, our findings align with those reported by~\textcite{uematsu2020}, where both $\Umod$ and $\Udom$, referred to as $\text{A}^*$-MOD and $\text{A}^*$-DOM, were evaluated for \textsc{Submodular Cardinality-Constrained Maximization}.

%%%%%%%%%%%%%%%%%%%%%%%%%%%%%%%%%%%%%%%%%%%%%%%%%%%%%%%%%%%%%%%%%%%%%%%%%%%%%%%%%%%%%%%%%%%%%%%%%%%%%%%%%%%%%%%%%%%%%%%%%%%%%%%%%%%%%%%%%%%%%%%%%%%%%%%%%%%%%%%%%%%%%%%%%%%%%%%%%%%%%%%%%%%%%%%%%%%%%%%%%%%%%%%%%%%%%%%%%%%%%%%%%%%%%%

\section{Discussion.}\label{e_sec5}
We presented a new branch-and-bound algorithm (Algorithm~\ref{e_alg1}) for solving the \textsc{Submodular Knapsack-Constrained Maximization} problem and several acceleration techniques for this basic algorithm. 
As our experimental results show, our methods are highly effective, particularly $\LECREP$, the combination of the acceleration techniques Lazy Evaluations with the average decision rule, Candidate Reduction, and Early Pruning. This combined approach strictly outperforms the solvers $\Umod$ and $\Udom$ by~\textcite{sakaue2018} across all instances of all problem types as well as the branch-and-cut algorithm, except for the $\cov$-instance based on real-world data.

A promising direction for future research is to reduce the running time of the proposed algorithms by parallelizing them. Given an instance of the \textsc{Submodular Knapsack-Constrained Maximization} problem and a fixed node in a basic search tree, the calls to $\sn$ for the children of that node, as they appear in the proposed algorithms, can be executed in parallel. The results from the subtrees rooted at these child nodes can then be used to select the optimal solution for the instance.

Another direction is to further enhance the efficiency of the pruning rule. For example, alternative decision functions for the lazy evaluation of a search tree beyond those presented could be considered. Specifically, instance-specific decision functions may improve performance.

\section*{Acknowledgements.} The authors thank the DFG for their support within RTG 2126 ``Algorithmic Optimization''.

\begingroup
  \renewcommand{\bibname}{References}
  \makeatletter
  \renewcommand{\@mkboth}[2]{} % deaktiviert das automatische \markboth von bibtex
  \makeatother
  \let\chapter\section % verhindere Seitenumbruch
  \printbibliography
\endgroup

\begin{subappendices}
\appendix
\renewcommand{\thesection}{\Alph{section}}

\section{Appendix: \texorpdfstring{Detailed discussion of the algorithm $\Udom$.}{}}
For \textsc{Submodular Knapsack-Constrained Maximization},~\textcite{sakaue2018} introduced the solver $\Udom$. In the following, we discuss possible implementations of $\Udom$ and address contradictory descriptions of $\Udom$ by~\textcite{sakaue2018}.

As mentioned, $\Umod$ and $\Udom$ are both best-first search algorithms that rely on heuristic functions to decide which node of a basic search tree should be considered next. 
Given an instance $(I,f,w,B)\in \I$ and a basic search tree $T=(V,A)$,~\textcite{sakaue2018} assume that any $c\in C(S)$ satisfies $w_c\leq B-w(S)$, since otherwise $S\cup \{c\}$ is not a feasible solution. In the following discussion, we adopt this assumption.
We now restate some definitions and algorithms by~\textcite{sakaue2018} to clarify how the heuristic function $u_{\text{dom}}$ used in $\Udom$ is defined and to highlight the issues associated with these definitions.

For any node $S\in V$ and $Y\subseteq C(S)$,~\textcite{sakaue2018} define
\begin{align*}
   u_{\text{mod}}(Y)\coloneqq \max\lbrace &\sum_{x\in C(S)\setminus Y} p_x f(x\vert S\cup Y) \colon \sum_{x\in C(S)\setminus Y}p_x w(x)\leq B-w(S),\\
   & p_x\in [0,1] \text{ for }c\in C(S)\setminus Y \rbrace,
\end{align*}
which provides an upper bound on $\max_{X\subseteq C(S)\setminus Y, w(X)\leq B-w(S)} f(X\vert S\cup Y)$.
Notice that $u_{\text{mod}}$ implicitly depends on a given node $S \in V$, even though this is not reflected in its notation.

Furthermore, ~\textcite{sakaue2018} define the heuristic $u_{\text{dom}}$ based on the following greedy algorithm.

\begin{algorithm}[H]
\DontPrintSemicolon
\caption{}\label{e_alg5}
\KwIn{$(I,f,w,B)\in \I$ and a node $S$ in a basic search tree.}
\KwOut{$Z\subseteq C(S)$.}
Choose $c_{\text{max}}\in \arg \max_{c\in C(S)} f(c\vert S)$\;
$X \leftarrow \emptyset$\;
\While{$C(S)\neq \emptyset$}{
Choose $c^* \in \arg \max_{c\in C(S)}\left\{\frac{f(c\vert S\cup X)}{w_c}\right\}$\;
\lIf{$w(X\cup \{c^*\})\leq B-w(S)$}{$X \leftarrow X\cup \{c^*\}$\label{e_alg5l5}
}
$C(S)\leftarrow C(S)\setminus \{c^*\}$\label{e_alg5l6}
} 
$Z\leftarrow \arg \max_{r\in\{c_{\text{max}},X\}} f(r\vert S)$\;\label{e_alg5l7}
\KwRet{Z}
\end{algorithm}

Assume that Algorithm~\ref{e_alg5} is applied to an instance $(I,f,w,B)\in \I$ and a node $S$ in a corresponding basic search tree.
Then,~\textcite{sakaue2018} use $X_{1:k}$ to denote the set $X$ constructed in the while loop in Algorithm~\ref{e_alg5}. Further, they use $X_{1:i}$ with $1\leq i\leq k$ to denote the set containing the first $i$ elements that are added in Line~\ref{e_alg5l5} of Algorithm~\ref{e_alg5} to $X$. For $i<1$, $X_{1:i}$ is defined as $\emptyset$.

For any $i\leq k$, let $c_i$ be the $i$-th element added in Line~\ref{e_alg5l5} of Algorithm~\ref{e_alg5} to $X$. Then, for any $i\leq k$,~\textcite{sakaue2018} define 
\begin{align*}
    \beta_i\coloneqq1-\frac{f(c_i\vert X_{1:i-1}\cup S)}{u_{\text{mod}}(X_{1:i-1})},\quad   \beta_{1:k}\coloneqq\begin{cases}
        0 &\text{if } u_{\text{mod}}(X_{1:i})= 0 \text{ for some } i\leq k, \\
        \prod_{i=1}^{k} \beta_i  &\text{otherwise.}
    \end{cases}
\end{align*}

The heuristic function $u_{\text{dom}}$ in a node $S$ is defined by~\textcite{sakaue2018} as 
\begin{align*}
    u_{\text{dom}}\coloneqq \frac{f(X_{1:k}\vert S)}{1-\beta_{1:k}}.
\end{align*}

Note that $u_{\text{dom}}$ always implicitly depends on a node $S$, even though this dependency is not reflected in the notation.
\textcite{sakaue2018} state that, for a given node $S$, the value $u_{\text{dom}}$ can be computed in parallel with a single execution of Algorithm~\ref{e_alg5} for $(I,f,w,B)$ and $S$, since all sets $X_{1:i-1}$ and all marginal gains $f(c\vert S\cup X_{1:i-1})$ for $c\in C(S)\setminus X_{1:i-1}$ to compute $u_{\text{mod}}(X_{1:i-1})$, and thus $u_{\text{dom}}$, are already computed during the execution of Algorithm~\ref{e_alg5}.
However, to calculate $u_{\text{mod}}(X_{1:i-1})$ for some $i\leq k$, and thus to compute $u_{\text{dom}}$, it is generally necessary to compute the marginal gain $f(c\vert S\cup X_{1:i-1})$ of an element $c$ that was not added to $X$ in Line~\ref{e_alg5l5}, but was removed from $C(S)$ in Line~\ref{e_alg5l6} of Algorithm~\ref{e_alg5}, in some iteration $j< i-1$ of the while loop.

This reveals a contradiction between the definitions given by~\textcite{sakaue2018} and the description provided to compute $u_{\text{dom}}$ for a node $S$.
To resolve this contradiction and avoid additional calculations, we implemented $\Udom$ using the following variant of a greedy algorithm.

\begin{algorithm}[H] 
%\SetAlgoLined
\DontPrintSemicolon
\caption{}\label{e_alg6}
\KwIn{$(I,f,w,B)\in \I$ and a node $S$ in a basic search tree.}
\KwOut{$Z\subseteq C(S)$.}
Choose $c_{\text{max}}\in \arg \max_{c\in C(S)} f(c\vert S)$\;
$X \leftarrow \emptyset$\;
\While{$C(S)\neq \emptyset$}{
Choose $c^* \in \arg \max_{c\in C(S)\setminus X}\left\{\frac{f(c\vert S\cup X)}{w_c}\right\}$\;
\lIf{$w(X\cup \{c^*\})\leq B-w(S)$}{$X \leftarrow X\cup \{c^*\}$
}
\lElse{\textbf{break}}
}
$Z\leftarrow \arg \max_{r\in\{c_{\text{max}},X\}} f(r\vert S)$\;
\KwRet{Z}
\end{algorithm}

By using Algorithm~\ref{e_alg6}, we can compute $u_{\text{dom}}$ for a node $S$ as described by~\textcite{sakaue2018} without additional function evaluations, since Algorithm~\ref{e_alg6} terminates as soon as adding an element $c^*$ to $X$ would exceed the knapsack capacity $B-w(S)$.

\vspace{0.5\baselineskip}Although we cannot verify that our implementation of $\Udom$ via Algorithm~\ref{e_alg6} matches the original, we observed that it runs faster than with Algorithm~\ref{e_alg5}. For this reason, we used it to compare against the algorithms proposed by us.

\vspace{3\baselineskip}

\section{\texorpdfstring{Appendix: Empirical Results for $\LEgCR$, $\LEgEP$, and $\LEgCREP$.}{}}

\vspace{2\baselineskip}

\begin{table}[H]
\centering
\begingroup
\renewcommand{\arraystretch}{0.93}
\begin{tabular}{C{1.cm} C{2.7cm} | C{2.5cm} C{2.5cm} C{3cm}}
\hline
&    & solved & time (s) & nodes  \\ \hline
\multirow{3}{*}{$\cov$} & $\LEgCR$  & $99$ & $176.69$ & $7342859$   \\
& $\LEgEP$    & $99$ & $187.02$ & $83701788$   \\
& $\LEgCREP$ & $99$ & $175.64$ & $7388209$   \\
\hline
\multirow{3}{*}{$\loc$} & $\LEgCR$  & $100$ & $74.66$ & $5575096$   \\
& $\LEgEP$    & $100$ & $86.72$ & $6805579$   \\
& $\LEgCREP$  & $100$ & $57.99$ & $5600549$   \\
\hline
\multirow{3}{*}{$\infe$}& $\LEgCR$ & $100$ & $21.29$ & $591112$  \\
& $\LEgEP$   & $100$ & $39.07$ & $1270962$   \\
& $\LEgCREP$ & $100$ & $21.51$ & $591166$   \\
\hline
\end{tabular}
\endgroup
\captionsetup{font=small}
\caption{Number of solved instances, average computation time (s), and average number of processed nodes for all solvers on artificial $\cov$-, $\loc$-, and $\infe$-instances generated following the method by~\textcite{sakaue2018}.}
\end{table}

\begin{table}[H]
\centering
\begingroup
\renewcommand{\arraystretch}{0.93}
\begin{tabular}{C{1.cm} C{2.7cm} | C{2.5cm} C{2.5cm} C{3cm}}
\hline
&    & solved & time (s) & nodes  \\ \hline
\multirow{3}{*}{$\cov$} & $\LEgCR$  & $1$ & $2786.32$ & $363420364.0$\\
& $\LEgEP$    & $1$ & $3208.85$ & $834320179.0$\\
& $\LEgCREP$  & $1$ & $2063.58$ & $363420364.0$\\
\hline
\multirow{3}{*}{$\loc$} & $\LEgCR$  & / & / & /   \\
& $\LEgEP$    & / & / & /   \\
& $\LEgCREP$  & / & / & /   \\
\hline
\multirow{3}{*}{$\infe$}& $\LEgCR$ & $2$ & $1630.6$ & $75419487$\\
& $\LEgEP$   & $1$ & $1758.37$ & $109731524$\\
& $\LEgCREP$ & $2$ & $1479.39$ & $75419487$ \\
\hline
\end{tabular}
\endgroup
\captionsetup{font=small}
\caption{Number of solved instances, average computation time (s), and average number of processed nodes for all solvers on artificial $\cov$-, $\loc$-, and $\infe$-instances generated following our method.}
\end{table}

\begin{table}[H]
\centering
\begin{tabular}{C{1.cm} C{2.7cm} | C{2.5cm} C{3cm}}
\hline
&   & time (s) & nodes  \\ \hline
\multirow{3}{*}{$\cov$} & $\LEgCR$ & $3016.53$ & $132044065$ \\
& $\LEgEP$   &  $3583.6$ & $1073706857$  \\
& $\LEgCREP$ &  $1275.12$ & $132044070$   \\
\hline
\end{tabular}
\captionsetup{font=small}
\caption{Computation time (in seconds) and number of processed nodes for the $\cov$-instance based on
real-world data}
\end{table}

None of the solvers $\LEgCR$, $\LEgEP$, and $\LEgCREP$ succeeded in solving the $\infe$-instance based on real-world data.

\end{subappendices}
%---------------------------------------------------------------------------------------------------------------------	
\end{refsection}

%---------------------------------------------------------------------------------------------------------------------	
%---------------------------------------------------------------------------------------------------------------------	
%---------------------------------------------------------------------------------------------------------------------	

\end{document}